\documentclass[12pt]{iopart}

\usepackage{epsfig}
\usepackage{graphicx}
\usepackage{dcolumn}
\usepackage{bm}
\usepackage[english]{babel}
\usepackage{iopams}
\usepackage{epstopdf}

\begin{document}

\title[Self-Consistent Description of Local Density Dynamics in Simple Liquids]{Self-Consistent Description of Local Density Dynamics in Simple Liquids. \\ The Case of Molten
Lithium}

\author{A V Mokshin and B N Galimzyanov}
\address{Department of Computational Physics, Institute of
Physics, Kazan Federal University, 420008 Kazan, Russia}

\ead{anatolii.mokshin@mail.ru}

\date{\today}

\begin{abstract}
The dynamic structure factor is the quantity, which can be measured by means of
Brillouin light-scattering as well as by means of inelastic
scattering of neutrons and X-rays. The spectral (or frequency)
moments of the dynamic structure factor define directly the sum
rules of the scattering law. The theoretical scheme formulated in
this study allows one to describe the dynamics of local density
fluctuations in simple liquids and to obtain the expression of the
dynamic structure factor in terms of the spectral moments. The
theory satisfies all the sum rules, and the obtained expression
for the dynamic structure factor yields correct extrapolations
into the hydrodynamic limit as well as into the free-particle
dynamics limit. We discuss correspondence of this theory with the
generalized hydrodynamics and with the viscoelastic
models, which are commonly used to analyze the data of inelastic neutron and X-ray scattering in liquids. In particular, we reveal that the postulated condition of the viscoelastic model for the memory function can be directly obtained within the presented theory.  The dynamic structure factor of liquid lithium is computed on the basis of the presented theory, and various features of the scattering spectra are evaluated. It is found that the theoretical results are in agreement with
inelastic X-ray scattering data.
\end{abstract}


\vspace{1.0cm} \noindent{\it Keywords\/}: microscopic dynamics,
liquids, liquid metals, density fluctuations, scattering spectra,
spectral moments, molecular dynamics simulations

\maketitle

\section{Introduction \label{sec: intr}}

The collective dynamics in a liquid occurs over a wide range of
the spatial scales varied from the scales comparable
with the particle sizes (of the order of few Angstroms) up to the
macroscopical scales typical for hydrodynamics. The features of
this dynamics determine directly the lineshape of the dynamic
structure factor $S(k,\omega)$. The long-wavelength limit ($k \to
0$) corresponds to ordinary hydrodynamics, and the spectrum of the
dynamic structure factor $S(k,\omega)$ at extremely small values
of the wavenumber $k$ is characterized by the line shape known as
the Rayleigh-Mandelshtam-Brillouin triplet. This spectrum consists of one central (Rayleigh) peak located at $\omega=0$ and two side (Brillouin) peaks located at frequencies $\pm \omega_c$, where $\omega_c = c_s k$ and $c_s$ is
the adiabatic sound velocity. Hence, the lineshape of
$S(k,\omega)$ represents a result of the sum of three Lorentzians (in
respect to the frequency $\omega$)~\cite{Resibua}:
\begin{eqnarray}\label{eq: hydrodynamics}
  S(k,\omega) &=& \frac{S(k)}{2 \pi} \left \{  \left ( \frac{\gamma - 1}{\gamma}  \right ) \frac{2 D_T k^2}{\omega^2 + ( D_T k^2)^2}  \right . \\
   & & \left .  +\frac{1}{\gamma} \sum_{l=1}^2\frac{\Gamma k^2}{[\omega+ (-1)^l c_s k]^2 + (\Gamma k^2)^2}    \right \}, \nonumber
\end{eqnarray}
with the following parameters -- the thermal diffusivity $D_T$,
the sound attenuation coefficient $\Gamma = (1/2)[(\gamma-1)D_T +
\eta_L]$, the ratio of the specific heat at constant pressure to
the specific heat at constant volume $\gamma = c_p/c_V$, the
longitudinal viscosity $\eta_L$, and the static structure factor
$S(k)$. Note that Eq.~(\ref{eq: hydrodynamics}) reproduces the next peculiarities of the density fluctuations dynamics observed in the Brillouin light-scattering experiments: (i) the
three peaks corresponding to these Lorentzians are well defined
and separated by two gaps of the same width; the Rayleigh
component $I_R$ and the Brillouin component $I_B$ are defined as
follows
 \begin{eqnarray}
I_R &=& \frac{ S(k)}{\pi} \left ( \frac{\gamma - 1}{\gamma} \right ) \int_{-\infty}^{\infty} \frac{D_T k^2}{ \omega^2 + (D_T k^2)^2} d\omega \nonumber \\
    &=& S(k) \left ( \frac{\gamma - 1}{\gamma}  \right )  \nonumber
\end{eqnarray}
and
\begin{equation}
I_B = \frac{ S(k)}{2 \pi} \frac{ 1}{\gamma}  \int_{-\infty}^{\infty} \frac{\Gamma k^2}{ [\omega \pm (c_s k ) ]^2 + (\Gamma k^2)^2} d\omega = \frac{S(k)}{2} \frac{1}{\gamma}. \nonumber
\end{equation}
The correspondence between the components is defined by the Landau-Placzek ratio~\cite{Landau/Placzek_1934}
\begin{equation}\label{eq: Landau_Placzek_ratio}
  \frac{I_R}{2 I_B} = \gamma - 1; \nonumber
\end{equation}
(ii) these peaks are broaden with increase of the wavenumber in accordance with  $k^2$-dependence; and (iii) shift of two symmetric peaks to
higher frequencies with increase of the wavenumber $k$ defines the
linear low-$k$ asymptotic of the longitudinal sound
dispersion, $\lim_{k\to 0}\omega_c(k) = c_s k$.

The Brillouin light-scattering in liquids probes the
macroscopical longitudinal density fluctuations. Then, Inelastic
Neutron Scattering (INS) and Inelastic X-ray Scattering (IXS)
experiments allow one to extract an information about the local
density fluctuations corresponding to the microscopical  spatial
scales with the wavelengths raised by collective dynamics of few
amount of particles up to the wavelengths corresponding to
free-particle dynamics regime~\cite{D_Price_PRep}. Let $k_m$ be
the wavenumber corresponding to the first (principal) maximum of
the static structure factor $S(k)$. For the case of monatomic
liquids, the maximum of the static structure factor, $S(k_m)$, is
the largest one, and the quantity $r_m=2\pi/k_m$ will provide an
approximate estimate of an equilibrium distance between the centers
of two neighboring atoms. The wavenumbers available via INS and
IXS experiments cover a wide range, which includes  $k \in [0.1;
\; 5] \; k_m$. These experiments have revealed that the three-peak
lineshape of $S(k,\omega)$-spectra similar to the
Rayleigh-Mandelshtam-Brillouin triplet appears also at finite
wavenumbers till the boundary of the first Brillouin pseudo-zone
with $k=k_m/2$, albeit these peaks are not separated, and the
experimental $S(k,\omega)$-spectra are not reproducible by
Eq.~(\ref{eq: hydrodynamics})~\cite{Verkerk_2001,Copley_Lovesey}.
With further increase of the wavenumber $k$, the side peaks of the
dynamic structure factor spectra start to shift to the lower frequencies and these peaks disappear completely at $k \simeq
0.75\; k_m$. At these wavenumbers, the tangent to the dispersion curve
$\omega_c(k)$ takes a negative slope and the intensity of the central
spectral component, $S(k,\omega=0)$, grows.  Such changes of the scattering
spectra occur until the wave number approaches the value $k_m$, at
which the spectrum of the dynamic structure factor will be
represented by the central Rayleigh component. Thus, at the wavenumbers
comparable with $k_m$, the $S(k,\omega)$ spectrum covers a narrow
frequency range. This is known as the de Gennes narrowing
effect~\cite{deGennes,Montfrooij}. Further, the  wavenumbers $k \geq
k_m$ correspond to the transition range in the regime of free-particle
dynamics. Thus, at very large wavenumbers $k > k_m$, the characteristic wavelenths
are shorter than an interparticle distance.  Here, the dynamic structure
factor $S(k,\omega)$ is reproduced by the Gaussian function
\begin{equation}\label{eq: Gaussian}
  S(k,\omega) = \sqrt{\frac{1}{2\pi}} \frac{1}{v_T k} \exp\left( - \frac{\omega^2}{2 (v_T k)^2}  \right ),
\end{equation}
where $v_T$ is  particle velocity.

INS and IXS spectra for the extended wavenumber range $k \in
[0.1;\; 1]k_m$ provide a useful information to develop and to
test the theoretical models of local density fluctuations dynamics
in liquids. The theoretical analysis of these experimental data is usually performed within the time correlation functions (TCF's) formalism~\cite{Hansen/McDonald_book_2006} and on
the basis of ideas of the generalized hydrodynamics~\cite{Balucani}. At the present time, there are
various theoretical models of the dynamic structure factor $S(k,\omega)$ of simple liquids (for details see
reviews~\cite{Scopigno_RMP_2005,Pilgrim_review,Verkerk_2001}).
And, as it turned out, it is not difficult to develop a
theoretical scheme with a certain amount of adjustable parameters.
Nevertheless, the main difficulty here is to propose a theoretical methodology
that will ensure the correct transition from hydrodynamic
description with a set of macroscopic parameters to description of
microscopic dynamics with appropriate characteristics such as the
interparticle interaction, the particle velocities and the
particle distribution functions. Moreover, a suggested theoretical model has to provide the correct time dependence of the TCF's and has to satisfy the so called sum-rules of a spectral functions~\cite{Egelstaff1}.

In this work, we will demonstrate that  description of density fluctuations dynamics as well as analysis of INS and IXS data of simple liquids can be done self-consistently in terms of the low-order frequency moments (or the low-order frequency parameters) of the dynamic structure factor. In fact, this theoretical scheme formalizes the idea of a self-consistent description of the density fluctuations in the same manner as this is realizing by the mode-coupling theory, where the statistical treatment of the dynamics of a system with an infinite number of degrees of freedom is reduced to closed integro-differential equations~\cite{Gotze_1989}. As follows from the theoretical definitions, the frequency moments are related to the microscopic characteristics: the particle interaction energy, the particle velocity, the  local structure parameters. This theoretical scheme allows one to obtain the model for
the dynamic structure factor of simple liquids.

Liquid alkali metals are suitable candidates to test the proposed
theories of the microscopical collective dynamics in a liquid~\cite{Egelstaff1,Boon/Yip_1991,Hansen/McDonald_book_2006,Balucani}.
The potential energy of these systems can be regarded as a quantity
determined mainly by the ion-ion interactions, whilst the electron
screening effects are taken into account effectively. The
interactions between the particles (ions) in alkali metals are
reproduced by a potential $u(\mathbf{r})$, which is the
simplest in comparison with the potentials for other metals.
Therefore, various characteristics of the density fluctuation
dynamics in liquid alkali metals can be evaluated on the basis of
the proposed microscopic theories, where the potential is the main
input parameter. Note that experimental data
for liquid alkali metals~\cite{Ohse_1985}, including recent
results of Inelastic Neutron and X-ray Scattering
\cite{Scopigno_RMP_2005,Pilgrim_review,D_Price_PRep,Sinn1,Novikov_K,CopleyRowe,Bodensteiner,Monaco_Cs,Cabrillo_K},
provides a reliable basis for theoretical and numerical simulation
studies~\cite{Torcini_1995,Casas_Li,Canales_Li_1994,Gonz_1994,Yulmetyev1,Yulmetyev2,Yulmetyev3,Mokshin_JCP_2004,Mokshin_JETP_2006,Anta,Mokshin_JPCM_2006,Tankeshwar_2007}.

Liquid lithium has the simple electronic structure $1s^2 2s$ and is characterized by a large ratio of valence electrons to core electrons. This ratio is larger even than for other liquid alkali metals. As a result, some physical properties of lithium determined by the microscopic dynamics turn out to be specific. In particular, liquid lithium has the largest sound velocity and the smallest heat capacity in comparison with other alkalis. Namely, the sound velocity is $c_s = 4554$~m/s and the molar heat capacity at constant pressure is $c_p = 30.33$~J/(mol$\cdot$K), whereas the ratio of specific heats $\gamma = c_p/c_V = 1.065$ at the melting temperature $T_m=453$~K and ambient pressure~\cite{Ohse_1985}. The local and non-local pseudopotentials as well as the more sophisticated EAM and modified EAM (MEAM) potentials have been proposed for the case of liquid lithium. To examine these potentials, the structure and dynamical properties had been evaluated in works of Kresse~\cite{Kresse}, Canales et al.~\cite{Canales_Li_1994}, Torcini et al.~\cite{Torcini_1995}, Gonzalez et al.~\cite{Gonz_2001}, Anta and Madden~\cite{Anta}, Salmon et al.~\cite{Salmon} and others.  It was found that the spectra of the dynamic structure factor  obtained by means of molecular dynamics with a pair pseudopotential  have better agreement with IXS data than simulation results with the EAM potentials~\cite{Gonz_2001,Khusn_2018}. In addition, many various theoretical models of the dynamic structure factor were tested with the high quality IXS data for liquid lithium~\cite{Torcini_1995,Scopigno_Li1,Tankeshwar_2003}. Therefore,  we examine our theoretical scheme for the case of liquid lithium.

The paper is organized as follows. Theoretical description is given in Section~\ref{sec: theory}. Namely, the basic points and definitions of the time correlation functions formalism  are presented in Subsection~\ref{sec: fundamentals}. The self-consistent theory of the density fluctuations dynamics in simple liquids developed in the framework of this formalism is presented in Subsection~\ref{sec: formalism}. In this Subsection, we present the original theoretical results related with the detailed derivation of the expressions for the dynamic structure factor, dispersion relation for the longitudinal collective excitations and other spectral characteristics. Comparison with other theoretical schemes is given in Section~\ref{sec:
Correspondence}. We discuss the correspondence of our theoretical scheme with the known simple and extended viscoelastic models (Subsection~\ref{sec: ex_viscoel_model}), with the ordinary and generalized hydrodynamic theories (Subsections~\ref{sec: hydrodynamics} and \ref{sec: gen_hydro}) and with the model for the free-particle dynamics (Subsection~\ref{sec: free_part_limit}).  In section~\ref{sec: lithium}, the theoretical model is applied to compute the dynamic structure factor of liquid lithium and the theoretical results are compared with experimental IXS data for this liquid metal. The concluding remarks
are given in Section~\ref{sec: concl}.

\section{Theoretical description \label{sec: theory}}

\subsection{Fundamental notions \label{sec: fundamentals}}

According to statistical mechanics, if one treats an equilibrium liquid as a many-particle system, then it is convenient to apply the mathematical formalism of the correlation functions, distribution functions, the moments and cumulants of these functions. In particular, the time correlation functions allow us to describe the relaxation processes occurring in the system at different spatial scales~\cite{Zwanzig_book_2001}. As a result, the corresponding microscopic theory can be developed. In the given study, we are focused on the collective particle dynamics of a liquid. Therefore, we assume that the particle interaction potential and such the structural parameters as the particle distribution functions and the static structure factor are defined and can be used as input parameters of the theoretical scheme.

Let us consider an isotropic  system, which consists on $N$
classical particles of a mass $m$ enclosed in a volume $V$. As an initial dynamical variable we take the Fourier-component of the local density fluctuations
\begin{equation}\label{eq: density_fluct}
  \rho_{k}=\frac{1}{\sqrt{N}}\sum_{j=1}^N e^{ i\mathbf{k} \cdot
\mathbf{r}_{j} }, \ \ \ k = |\mathbf{k}|,
\end{equation}
whose evolution is defined by the corresponding equation of motion
\begin{equation}\label{eq: eq_motion}
  \dot{\rho}_{k}(t) = i \hat{\mathcal{L}}  \rho_{k}(t).
\end{equation}
Here, the dot denotes time differentiation, and $
\hat{\mathcal{L}} $ is the Liouville operator, which is
Hermitian~\cite{Zwanzig_book_2001,Lee3}:
\begin{equation}\label{eq: Liouvillian}
  \hat{\mathcal{L}} = -i \left \{ \frac{1}{m}\sum_{j=1}^{N} (\mathbf{p}_j \cdot \mathbf{\nabla}_j)  - \sum_{l>j=1}^{N} \nabla_j u(r_{jl}) \left ( \nabla_{\mathbf{p}_j} - \nabla_{\mathbf{p}_l} \right )
    \right \}.
\end{equation}
Further, $\mathbf{p}_j$ is the momentum of the $j$th particle, $\mathbf{\nabla}_j$
and $\nabla_{\mathbf{p}_j}$ are the gradients over coordinates and
momenta, respectively. The quantity $u(r)$ is the interaction energy between a pair of the particles and, as expected, it can be evaluated from any model potential. By means of the Gram-Schmidt
orthogonalization procedure~\cite{Lee1} on the basis of the quantity $A_0(k)
\equiv \rho_{k}$ we generate the infinite set of variables
\begin{equation} \label{eq: inf_set}
  \mathbf{A}(k) = \{ A_0(k),\; A_1(k),\; A_2(k),\; \ldots  \} \\
\end{equation}
related by
\begin{eqnarray} \label{eq: variables}
  A_{j+1}(k) = i \hat{\mathcal{L}} A_j(k) + \Delta_j^2(k) A_{j-1}(k), \\
  j = 0,\; 1,\; 2,\; \ldots ;\nonumber\\
  A_{-1} \equiv 0.\nonumber
\end{eqnarray}
Here,
\begin{equation} \label{eq: frequency_parameter}
 \Delta_{j+1}^2(k) = \frac{\langle |A_{j+1}(k)|^2 \rangle}{\langle |A_j(k)|^2 \rangle}
\end{equation}
is an $j$th-order frequency parameter (at fixed $k$), which has a
dimension of square frequency~\cite{Mokshin_TMF_2015}; and the
brackets $\langle \ldots \rangle $ denote the ensemble average.

The variables of set~(\ref{eq: inf_set}) are the generalized
dynamical variables dependent on the wavenumber $k$ as parameter~(see discussion in Ref.~\cite{Gotze_book_2009},
pages~100-101). These variables form an orthogonal
basis~\cite{Lee1,Lee3}:
\begin{eqnarray} \label{eq: ort}
  \langle A^*_i(k) A_j(k) \rangle = \delta_{i,j} \langle |A_j(k)|^2 \rangle,\\
  i,j=0,\;1,\;2,\;\ldots . \nonumber
\end{eqnarray}
Further, while the first variable $A_0(k)$ is defined by Eq.~(\ref{eq: density_fluct}), then the second variable is
\begin{equation}\label{eq: A1}
  A_1(k) = \frac{i}{\sqrt{N}} \sum_{j=1}^N (\mathbf{k} \cdot \mathbf{v}_j) e^{i\mathbf{k} \cdot
\mathbf{r}_{j}},
\end{equation}
and for the third variable one has
\begin{eqnarray} \label{eq: A2}
A_{2}(k) &=& \dot{A}_1(k) + \Delta_1^2(k) A_0(k) \\
&=&  - \frac{1}{\sqrt{N}} \sum_{j=1}^{N} 
\left( \mathbf{k} \cdot \mathbf{v}_{j} \right )^{2} \textrm{e}^{i
\mathbf{k} \cdot \mathbf{r}_{j}} \nonumber \\
&+& \frac{i }{\sqrt{mN}} \sum_{l>j=1}^{N} 
(\mathbf{\nabla}_j \cdot \mathbf{k}) u(r_{jl}) \left \{  \textrm{e}^{i
\mathbf{k} \cdot \mathbf{r}_j} - \textrm{e}^{i
\mathbf{k} \cdot \mathbf{r}_{l}} \right \} + \Delta_1^2(k) A_0(k). \nonumber
\end{eqnarray}
 Let us define the time correlation function (TCF)
\begin{eqnarray}\label{eq: TCF}
  M_{j}(k,t) = \frac{\langle  A_j^*(0) A_j(t) \rangle}{ \langle |A_j^*(0)|^2 \rangle},\\
   j = 0,\; 1,\; 2,\; \ldots, \nonumber
\end{eqnarray}
which will be characterized by the properties
\numparts
\label{eq: prop_TCF}
\begin{equation} \label{eq: prop1}
    \left . {M}_j(k,t)\right |_{t=0} = 1,
\end{equation}
\begin{equation}\label{eq: prop2}
  \left | M_j(k,t) \right | \leq 1,
\end{equation}
and
\begin{eqnarray}\label{eq: prop3}
        \left . \frac{d^{l}}{dt^l}M_{j}(k,t) \right |_{t=0} = 0,  & & \textrm{if \textit{l} is odd}, \\
        \left . \frac{d^{l}}{dt^l}M_{j}(k,t) \right |_{t=0} \neq 0, & &   \textrm{if \textit{l} is even}. \nonumber
\end{eqnarray}
\endnumparts
These properties are directly derived from condition~(\ref{eq: ort}). Then, the TCF
\begin{equation}\label{eq: TCF1}
  M_0(k,t) = F(k,t) = \frac{\langle \rho_k^*(0) \rho_k(t)\rangle}{\langle |\rho_k(0)|^2 \rangle}
\end{equation}
represents the density-density correlation function known also as the intermediate scattering function. This TCF is related to the dynamic structure factor~\cite{Hansen/McDonald_book_2006}
\begin{equation}\label{eq: dsf}
  S(k,\omega)=\frac{S(k)}{2\pi} \int_{-\infty}^{\infty} \; \exp(i\omega t) F(k,t)\; dt,
\end{equation}
where $S(k) = \langle | \rho_{k}(0) |^2 \rangle$ is the static structure factor.
Taking into account property~(\ref{eq: prop3}), one obtains the short-time expansion for the intermediate scattering function:
\[ \label{eq: Taylor_series}\nonumber
    F(k,t) = 1 - \langle \omega^{(2)}(k) \rangle \frac{t^2}{2!} + \langle \omega^{(4)}(k) \rangle \frac{t^4}{4!}
    + \ldots + (-i)^l \langle \omega^{(l)}(k) \rangle \frac{t^l}{l!} + \ldots,
\]
where $\langle \omega^{(l)} \rangle$ is the $l$th-order normalized
frequency moment of the dynamic structure factor
$S(k,\omega)$~\cite{Tkachenko_1,Tkachenko_2,ind,Ming,Wierling}:
\begin{equation}\label{eq: freq_moments}
  \langle \omega^{(l)}(k) \rangle = \left . (-i)^l \frac{d^{(l)}}{dt^l}F(k,t) \right |_{t=0}
  = \frac{\int_{-\infty}^{\infty} \omega^l S(k,\omega) d\omega }{\int_{-\infty}^{\infty} S(k,\omega) d\omega}. \nonumber
\end{equation}

The quantity $M_1(k,t)$ is the TCF
associated with the longitudinal-current autocorrelation function
\begin{equation}\label{eq: TCF_current}
  C_L(k,t) = \frac{\langle j_k^{(L)}(0) j_k^{(L)}(t)   \rangle}{\langle |j_k^{(L)}(0)|^2 \rangle},
\end{equation}
which is related, in turn, to the intermediate scattering function $F(k,t)$ as follows~\cite{Boon/Yip_1991}
\begin{equation}\label{eq: TCF1_TCF2}
  - \frac{d^2}{dt^2} F(k,t) = \Delta_1^2(k) C_L(k,t).
\end{equation}
The label $L$ denotes the component parallel to the wavevector
$\mathbf{k}$. Relation~(\ref{eq: TCF1_TCF2}) is direct consequence
of the \textit{hydrodynamic} continuity equation
\begin{equation}\label{eq: continuity}
  \frac{\partial}{\partial t} \rho_k(t) + i k j_k^{(L)} =0,
\end{equation}
which shows that spontaneous fluctuations of a conserved
\textit{hydrodynamic} variable $\rho_k(t)$ decay very slowly at
long wavelengths (i.e. at extremely small
wavenumbers)~\cite{Hansen/McDonald_book_2006}. Then, taking into
account Eq.~(\ref{eq: dsf}) one obtains the correspondence between
the spectral density of the longitudinal-current fluctuations
\[
C_L(k,\omega) = \frac{1}{2 \pi} \int_{-\infty}^{\infty} \;
\exp(i\omega t) C_L(k,t)\; dt
\]
and the dynamic structure factor
$S(k,\omega)$~\cite{Chtchelkatchev_1,Chtchelkatchev_2}:
\begin{equation}\label{eq: relation_with_dsf}
  S(k) \Delta_1^2(k)C_L(k,\omega) = \omega^2 S(k,\omega).
\end{equation}

Further, from Eq.~(\ref{eq: A2}) one can verify that the TCF
$M_2(k,t)$ is expressed in terms of the TCF's of energy or stress
tensor, of force and of density as well as in terms of the
corresponding cross-correlation functions. Then,  the TCF
$M_3(k,t)$ will contain a contribution related to the TCF of
energy current~\cite{Yulmetyev3}. Thus, the quantities $F(k,t)$,
$M_1(k,t)$, $M_2(k,t)$ and $M_3(k,t)$ are the
TCF's of the variables  arising also in the hydrodynamic conservation laws for the longitudinal collective dynamics. These TCF's
correspond to the concrete relaxation processes~\cite{Mokshin_PRL_2005}. The time scales of the relaxation processes can be
conveniently evaluated by
\begin{equation}\label{eq: time_scales}
  \tau_j(k) = \frac{1}{\sqrt{\Delta_j^2(k)}}, \; \; j=0,\; 1,\; 2,\;
  \ldots.
\end{equation}

The functions $F(k,t)$, $C_L(k,t)$, $M_2(k,t)$ and $M_3(k,t)$ obey the next kinetic integro-differential equations:
\numparts
  \begin{equation}\label{eq: GLE1}
    \ddot{F}(k,t)+\Delta_{1}^{2}(k)F(k,t)+ \Delta_{2}^{2}(k) \int_{0}^{t}d\tau M_{2}(k,t-\tau) \dot{F}(k,\tau^{'})=0,
  \end{equation}
  \begin{equation}\label{eq: GLE2}
    \ddot{M_1}(k,t)+\Delta_{2}^{2}(k)M_1(k,t)+ \Delta_{3}^{2}(k) \int_{0}^{t}d\tau M_{3}(k,t-\tau) \dot{M_1}(k,\tau^{'})=0,
  \end{equation}
  \begin{equation}\label{eq: GLE3}
    \ddot{M_2}(k,t)+\Delta_{3}^{2}(k)M_2(k,t)+ \Delta_{4}^{2}(k) \int_{0}^{t}d\tau M_{4}(k,t-\tau)
    \dot{M_2}(k,\tau^{'})=0.
  \end{equation}
\endnumparts
Eqs.~(\ref{eq: GLE1}), (\ref{eq: GLE2}) and (\ref{eq: GLE3}) can
be derived by means of the recurrent relations method~\cite{Lee1}
or by the projection operators technique~\cite{Zwanzig_book_2001}
from the equations of motion for the variables $A_0(k)$, $A_1(k)$
and $A_2(k)$. Recall that the frequency parameters $\Delta_1^2(k)$,
$\Delta_2^2(k)$, $\Delta_3^2(k)$ and $\Delta_4^2(k)$ can be found directly from
definition~(\ref{eq: frequency_parameter}). On the other hand, one obtains from (\ref{eq: freq_moments}) that
\numparts
\begin{equation}\label{eq: Delta1}
    \Delta_1^2(k) = \langle \omega^{(2)}(k) \rangle,
\end{equation}
\begin{equation}\label{eq: Delta2}
    \Delta_2^2(k) = \frac{\langle \omega^{(4)}(k) \rangle}{\langle \omega^{(2)}(k) \rangle} - \langle \omega^{(2)}(k) \rangle,
\end{equation}
\begin{equation}\label{eq: Delta3}
    \Delta_3^2(k) = \frac{\langle \omega^{(6)}(k) \rangle \langle \omega^{(2)}(k) \rangle - (\langle \omega^{(4)}(k) \rangle)^2}{\langle \omega^{(4)}(k) \rangle \langle \omega^{(2)}(k) \rangle - (\langle \omega^{(2)}(k) \rangle)^3},
\end{equation}
\begin{eqnarray}\label{eq: Delta4}
\Delta_4^2(k) &=& \frac{1}{\Delta_1^2(k)\Delta_2^2(k)\Delta_3^2(k)} \Big\{\langle \omega^{(8)}(k) \rangle - \Delta_1^2(k) \Big[\Big(\Delta_1^2(k)+\Delta_2^2(k)\Big)^3 + \nonumber\\ & & + 2\Delta_2^2(k)\Delta_3^2(k) \Big(\Delta_1^2(k)+ \Delta_2^2(k)\Big)  + \Delta_2^2(k)\Delta_3^4(k)
\Big]\Big\},
\end{eqnarray}
\begin{eqnarray} \label{eq: Delta5}
  \Delta_5^2(k) &=& \frac{1}{\Delta_{1}^{2}(k)\Delta_{2}^{2}(k)\Delta_{3}^{2}(k)\Delta_{4}^{2}(k)}\Bigg\{\omega^{(10)}(k)- \\ & & - 2\omega^{(8)}(k) \Big[\Delta_{1}^{2}(k)+\Delta_{2}^{2}(k)+ \Delta_{3}^{2}(k)+\Delta_{4}^{2}(k)\Big]+
\nonumber\\ & &
+\omega^{(6)}(k)\Big[\Delta_{1}^{4}(k)+2\Delta_{1}^{2}(k)\Delta_{2}^{2}(k)+\Delta_{2}^{4}(k)+4\Delta_{1}^{2}(k)\Delta_{3}^{2}(k)
\nonumber \\& &
+2\Delta_{2}^{2}(k)\Delta_{3}^{2}(k)+\Delta_{3}^{4}(k)+4\Delta_{1}^{2}(k)\Delta_{4}^{2}(k)
\Big]\Biggl\}+\Delta_{4}^{2}(k). \nonumber
\end{eqnarray}
\endnumparts
One can see that the frequency parameter of the $j$th order is defined
through the frequency moments up to the moment of the $2j$th
order. Relations similar to (\ref{eq: Delta1})--(\ref{eq:
Delta5}) can be also written for the frequency parameters of the
higher order. Moreother, relations~(\ref{eq: freq_moments}), (\ref{eq: Delta1})--(\ref{eq: Delta5}) are the
low-order sum rules of the scattering law $S(k,\omega)$.

Taking into account the statistical average in
the product of the dynamical variables in relation~(\ref{eq:
frequency_parameter}),  the frequency moments $\langle
\omega^{(l)}(k) \rangle$ can be expressed in terms of the thermal
velocity $v_{th} = \sqrt{k_BT/m}$ of a particle, the interparticle
potential $u(\mathbf{r})$ and the particles distribution
functions. Thus, the first frequency parameter $\Delta_1^2(k)$
is defined as
\numparts \label{eq: frequencies}
  \begin{equation}\label{eq: D1}
    \Delta_1^2(k) = \frac{k_B T}{m} \frac{k^2}{S(k)} = \frac{(v_{th} k)^2}{S(k)}.
  \end{equation}
The second frequency parameter $\Delta_2^2(k)$  can be found from
  \begin{eqnarray}\label{eq: D2}
    \Delta_2^2(k) 
                  &=& 3 \frac{k_B T}{m}k^2 - \Delta_1^2(k) + \frac{\rho}{m} \int \nabla_l^2 u(\mathbf{r})[1 - \exp(i\mathbf{k}\cdot \mathbf{r})]g(r)\mathrm{d}^3\mathbf{r}  \\
                  &\approx& 3 \left ( \frac{k_B T}{m}k^2 +  \omega_E^2 \right ) - \Delta_1^2(k) - \frac{\rho}{m} \int \nabla_l^2 u(\mathbf{r})\exp(i\mathbf{k}\cdot \mathbf{r})g(r)\mathrm{d}^3\mathbf{r} .
                  \nonumber
  \end{eqnarray}
  The third frequency parameter is defined as
  \begin{eqnarray}\label{eq: D3}
    \Delta_3^2(k) &=& \frac{1}{\Delta_2^2(k)} \left \{ 15 \left (  \frac{k_B T}{m}k^2 \right ) + \mathcal{F}(k)    \right \}  - \frac{1}{\Delta_2^2(k)} \left [    \Delta_1^2(k) + \Delta_2^2(k) \right ]^2.
  \end{eqnarray}
\endnumparts
Here, $\rho = N/V$ is the number density, $g(r)$ is the pair
distribution function,  $\omega_E$ is the so-called Einstein
frequency, and the term $\mathcal{F}(k)$ denotes the combination
integral expressions containing the interparticle potential
$u(\mathbf{r})$ and such the structural characteristics as the
two- and three-particle distribution
functions~\cite{Mokshin_JPCM_2006}. Moreover, the $j$th order relaxation parameter $\Delta_j^2(k)$ with
$j\geq 2$ will be defined by the potential $u(\mathbf{r})$ as well
as by the distribution functions of two, three, $\ldots$, $(j-2)$,
$(j-1)$ and $j$ particles. Thus, the relaxation parameters
$\Delta_j^2(k)$ at $j \geq 2$ are directly dependent on the type of
interaction between the particles and on the microscopic
structural characteristics.

Eq.~(\ref{eq: GLE1}) is also known as the generalized Langevin equation. By the Laplace transformation
\[
LT\{ f \}(s)= \tilde{f}(s) = \int_{0}^{\infty}
e^{-st}f(t)\; dt, \hskip 1cm s=i\omega,
\]
of Eq. (\ref{eq: GLE1}) and solving it
in terms of Eq. (\ref{eq: dsf}), one obtains expression
for the \textit{classical} dynamic structure factor:
\begin{equation} \label{eq: DHOg}
S(k,\omega)=\frac{S(k)}{\pi} \frac{\Delta_{1}^{2}(k)
\Delta_{2}^{2}(k)
M'_{2}(k,\omega)}{[\omega^{2}-\Delta_{1}^{2}(k)+\omega
\Delta_{2}^{2}(k) M''_{2}(k,\omega)]^{2}+[\omega \Delta_{2}^{2}(k)
M'_{2}(k,\omega)]^{2}},
\end{equation}
where
$\widetilde{M}'_{2}(k,\omega)$ and $\widetilde{M}''_{2}(k,\omega)$ are the real and imaginary parts of $\widetilde{M}_{2}(k,s=i\omega)$, respectively.
Eq.~(\ref{eq: DHOg}) is usually applied to explain the features of the experimental dynamic structure factor~\cite{Verkerk_2001}. This implies  in fact a fitting of the experimental $S(k,\omega)$ by Eq.~(\ref{eq: DHOg}) with a suggested theoretical model for the TCF $M_2(k,t)$~\cite{Hansen/McDonald_book_2006}.
In particular, if the TCF decays instantaneously, $M_2(k,t) = \tau_2(k) \delta(t)$, then Eq.~(\ref{eq: DHOg}) reduces to the equation for the \textit{damped harmonic oscillator} (DHO) model:
\begin{equation} \label{eq: DHO}
S(k,\omega)=\frac{S(k)}{\pi} \frac{\Delta_{1}^{2}(k)
\Delta_{2}^{2}(k)
\tau_2(k)}{[\omega^{2}-\Delta_{1}^{2}(k)]^{2}+[\omega \Delta_{2}^{2}(k)
\tau_2(k)]^{2}},
\end{equation}
where $\tau_2(k)$ is the damping parameter. On the other hand, the TCF with an exponential decay
\begin{equation} \label{eq: viscoel_M2}
M_2(k,t) = e^{- t/\tau_2(k)}
\end{equation}
yields the simple \textit{viscoelastic model} for the dynamic structure factor $S(k,\omega)$~\cite{Copley_Lovesey}.

\subsection{Self-Consistent approach \label{sec: formalism}}

The equations similar to kinetic Eqs.~(26) can
be written for the TCF's of other dynamical variables --
$M_3(k,t)$, $M_4(k,t)$, $\ldots$, $M_j(k,t)$, $\ldots$. As a result, one obtains the infinite chain of the connected
equations~\cite{Hansen/McDonald_book_2006}. Applying the Laplace
transform to these equations, one obtains a continued fraction
representation for the Laplace transform of the intermediate
scattering function $\widetilde{F}(k,s)$:
\begin{eqnarray} \label{eq: cont_fraction}
\widetilde{F}(k,s)&=&\frac{1}{s+\Delta_{1}^{2}(k)\widetilde{M}_{1}(k,s)} \\
&=&
\frac{1}{\displaystyle s+\frac{\Delta_{1}^{2}(k)}{\displaystyle
s+\frac{\Delta_{2}^{2}(k)}{\displaystyle
s+\frac{\Delta_{3}^{2}(k)}{\displaystyle s+\ddots}}}}.\nonumber
\end{eqnarray}
This fraction indicates that the dynamic structure factor
\begin{equation} \label{eq: Sko}
S(k,\omega) = \frac{S(k)}{\pi} \mathrm{Re} [ \widetilde{F}(k,s=i\omega)]
\end{equation}
is directly defined by the frequency parameters $\Delta_j^2(k)$'s
and, thereby, by the dynamical variables of set~(\ref{eq:
inf_set}). Estimation of the low-order frequency parameters for
liquid metals (cesium~\cite{Yulmetyev1,Yulmetyev2},
sodium~\cite{Mokshin_JCP_2004,Yulmetyev3},
aluminium~\cite{Tankeshwar_2007,Yulmetyev3,Mokshin_JETP_2006})
sets  the following regularity:
\begin{eqnarray}
\Delta_{j+1}^2(k) &\geq&  \Delta_{j}^2(k), \hskip 1cm j=1,\;
2,\;3,\; \ldots, \\
& &k \mathrm{\ is\ fixed}.\nonumber
\end{eqnarray}
This is evidence that the relaxation process associated
with a higher-order dynamical variable takes place on a shorter
time-scale~\cite{Mokshin_PRL_2005}.~\footnote{Increase of
parameters $\Delta_j^2(k)$ with growing $j$ is not surprising,
because the spectral moment (and, therefore, the corresponding
frequency parameter) of a higher order characterizes the spectral features with higher frequencies. Moreover, rigorous inequality
$\Delta_{j+1}^2(k) > \Delta_j^2(k)$ at $j=1$ is well-known for the
case of classical liquids (see, for example, on pages 330-331 in
review~\cite{Yoshido_review}).} Then, one
can reasonably expect for the high-order frequency
parameters starting from the parameter with a some index $\xi$ that
\begin{equation} \label{eq: cond}
{\Delta_{\xi}^2(k)} \gg {\omega^2}, \hskip 1 cm \xi \; \mathrm{is\ natural\ number},
\end{equation}
where $\omega$'s is the inherent frequencies of the relaxation
processes associated with the density fluctuations. These
frequencies $\omega$ correspond to the range  (at the given
$k$), where the inelastic components of $S(k,\omega)$
are observed. This means that the time scale of density
fluctuations $\tau_{\alpha}(k) \propto 1/\sqrt{\Delta_1^2(k)}$ is
much larger than the associated time scales
$\tau_{\xi}(k)=1/\sqrt{\Delta_{\xi}^2(k)}$,
$\tau_{\xi+1}(k)=1/\sqrt{\Delta_{\xi+1}^2(k)}$ and so on. Then,
assuming that the time scales take small but still finite values
one can write that
\begin{equation}\label{eq: time_scales_equality}
  \Delta_{\xi}^2(k) = \Delta_{\xi+1}^2(k) = \Delta_{\xi+2}^2(k) =\ldots.
\end{equation}
Condition~(\ref{eq: time_scales_equality}) has the important inferences.

\noindent (i) This condition corresponds exactly to the TCF $M_{\xi-1}(k,t)$ with the following time dependence~\cite{Mokshin_TMF_2015}:
\begin{equation}\label{eq: Mj_1}
  M_{\xi-1}(k,t) = \frac{1}{\sqrt{\Delta_{\xi}^2(k)}t} J_1 \left (2\sqrt{\Delta_{\xi}^2(k)}t\right ) .
\end{equation}
The Laplace transform of this TCF is
\begin{equation}\label{eq: LT_Mj_1}
  \widetilde{M}_{\xi-1}(k,s) = \frac{-s + [s^2 +4 \Delta_{\xi}^2(k)]^{1/2}}{2
  \Delta_{\xi}^2(k)}.
\end{equation}
Here, $J_1(x)$ is the Bessel function of the first kind. As a
result, the frequency dependence of all the quantities,
$\widetilde{F}(k,s)$, $\widetilde{M}_1(k,s)$,
$\widetilde{M}_2(k,s)$, $\ldots$, $\widetilde{M}_{\xi-2}(k,s)$,
can be recovered by substitution of (\ref{eq: LT_Mj_1}) into
fraction (\ref{eq: cont_fraction}).

\noindent
(ii) Condition~(\ref{eq: time_scales_equality}) is equivalent to equality of the TCF's~\cite{Mokshin_TMF_2015}:
\begin{equation}\label{eq: TCF_equality}
  M_{\xi}(k,t) = M_{\xi-1}(k,t),
\end{equation}
which yields the correspondence between the dynamical variables:
\begin{equation}\label{eq: dyn_var_prop}
  A_{\xi}(k) \sim A_{\xi-1}(k).
\end{equation}
The physical meaning of (\ref{eq: dyn_var_prop}) is that the set $\mathbf{A}(k)$ [see (\ref{eq: inf_set})] reduces to the finite amount of the dynamical variables:
\begin{equation} \label{eq: model_set}
A_0(k),\; A_1(k),\; A_2(k),\;\ldots,\; A_{\xi-1}(k).
\end{equation}
Relation~(\ref{eq: TCF_equality}) yields directly solution
of the chain of integro-differential equations~(26) in a
self-consistent way similar to that as this is realized in the
mode-coupling theories~\cite{Gotze_book_2009}.

To be consistent with hydrodynamic theory, the finite set of the
dynamical variables has to include, at least, first four dynamical
variables of set~(\ref{eq: inf_set}), i.e. $\xi=4$ in
set~(\ref{eq: model_set}), corresponding to the hydrodynamic variables~\cite{Hansen/McDonald_book_2006}. Then, [by analogy with
Eq.~(\ref{eq: LT_Mj_1})] one obtains the following $s$-dependence
of $\widetilde{M}_3(k,s)$:
\begin{equation}\label{eq: LT_M3}
  \widetilde{M}_{3}(k,s) = \frac{-s + [s^2 +4 \Delta_{4}^2(k)]^{1/2}}{2 \Delta_{4}^2(k)}.
\end{equation}
Further, the Laplace transform $\widetilde{M}_{2}(k,s) =
M'_2(k,\omega)+ i M''_2(k,\omega)$, which appears in general
expression~(\ref{eq: DHOg}) for the dynamic structure factor, takes the form
\begin{equation}\label{eq: LT_M2}
  \widetilde{M}_{2}(k,s) = \frac{2\Delta_4^2(k)}{s \left [ 2 \Delta_4^2(k) - \Delta_3^2(k) \right ] + \Delta_3^2(k) \sqrt{s^2+ 4\Delta_4^2(k)} },
\end{equation}
whereas the dynamic structure factor is
\numparts
\label{eq: Basic}
\begin{equation}
 \label{eq: Basic_a}   S(k, \omega)= \frac{S(k)}{2 \pi}
\frac{\Delta_{1}^2(k) \Delta_{2}^2(k)
\Delta_{3}^2(k)}{\Delta_4^2(k)-\Delta_3^2(k)} \frac{[4
\Delta_{4}^2(k)- \omega^{2}]^{1/2}}{\omega^6 +\mathcal{A}_1(k)
\omega^4 +\mathcal{A}_2(k) \omega^2 +\mathcal{A}_3(k)}
\end{equation}
with
\begin{eqnarray} \label{eq: coeff_A}
 \mathcal{A}_1(k)&=&
\frac{\Delta_3^4(k)-\Delta_2^2(k)[2\Delta_4^2(k)-\Delta_3^2(k)]}{\Delta_4^2(k)-\Delta_3^2(k)}
-2\Delta_1^2(k),\\
\mathcal{A}_2(k)&=&\frac{\Delta_2^4(k)\Delta_4^2(k)-2\Delta_1^2(k)\Delta_3^4(k)+\Delta_1^2(k)\Delta_2^2(k)
[2\Delta_4^2(k)-\Delta_3^2(k)]}{\Delta_4^2(k)-\Delta_3^2(k)}+\nonumber\\ & & + \Delta_1^4(k),\nonumber\\
\mathcal{A}_3(k)&=&\frac{\Delta_1^4(k)\Delta_3^4(k)}{\Delta_4^2(k)-\Delta_3^2(k)}.
\nonumber
\end{eqnarray}
\endnumparts

Taking into account the condition $4 \Delta_4^2(k) \gg \omega^2$,
which is equivalent to (\ref{eq: cond}) at $\xi=4$,
the numerator of (\ref{eq: Basic_a}) turns into  product of the
frequency parameters $\Delta_{1}^2(k)$,  $\Delta_{2}^2(k)$,
$\Delta_{3}^2(k)$ and $\Delta_4^2(k)$. Then, the spectral features
of the dynamic structure factor spectrum  $S(k,\omega)$ at a given
$k$ will be defined according to (\ref{eq: Basic_a}) by the
bicubic polynomial (in the variable $\omega$) in the denominator
with the coefficients $\mathcal{A}_1(k)$, $\mathcal{A}_2(k)$,
$\mathcal{A}_3(k)$. It should be noted that the scattering spectra defined
by Eq.~(\ref{eq: Basic_a}) satisfy all the sum rules.

The fact that the  higher order frequency parameters,
$\Delta_5^2(k)$, $\Delta_6^2(k)$, $\ldots$, are not included in
expression (\ref{eq: Basic_a}) for the dynamic structure factor is naturally due
to that the relaxation processes associated with such the
dynamical variables as the time derivatives of the energy current
have no direct influence on the longitudinal density fluctuations
dynamics. This is realized when the time scales of these
relaxation processes in a liquid are finite, comparable,
but  much shorter than the time scale of the density fluctuations
dynamics.  On the other hand, applying Eq. (\ref{eq: LT_Mj_1}) at
an arbitrary index $\xi
> 4$ to continued fraction (\ref{eq: cont_fraction}), expression
(\ref{eq: Basic_a}) is modified to the form, which includes
$\Delta_5^2(k)$, $\Delta_6^2(k)$,  etc. Nevertheless, if these
high-order frequency parameters have approximately equal values,
then the modified expression for the dynamic structure factor will
yield the same spectral lineshape as expression (\ref{eq: Basic_a}).

As seen from~(\ref{eq: Basic_a}), all the spectral features of the dynamic
structure factor $S(k,\omega)$ are directly defined by the
interaction potential $u(\mathbf{r})$ and by such the structural
characteristics as the distribution functions of two, three and
four particles. These quantities appear in expressions for the frequency
parameters $\Delta_1^2(k)$, $\Delta_2^2(k)$, $\Delta_3^2(k)$ and
$\Delta_4^2(k)$. The distribution functions of larger amount of particles do not appear in this theoretical model. Finally, it is necessary to note that no assumptions about amount of relaxation modes and about their
time/frequency dependencies were made to obtain expression~(\ref{eq: Basic_a}) for the dynamic structure factor.

Moreover, analysis of (\ref{eq: Basic_a}) reveals that the dynamic structure
factor $S(k,\omega)$ at fixed $k$ must be also characterized by three peaks located at \numparts \label{eq: maxima}
\begin{equation} \label{eq: central}
\omega_0=0 \ \ \ \ \ (\mathrm{central} \ \ \mathrm{peak})
\end{equation}
and
\begin{eqnarray} \label{eq: side}
 \omega^{(max)}_{+,-}(k) = \pm \sqrt{\frac{-\mathcal{A}_1(k)+\sqrt{\mathcal{A}_1(k)^2-3\mathcal{A}_2(k)}}{3}} \\
 (\mathrm{high-frequency} \ \ \mathrm{doublet}) \nonumber
\end{eqnarray}
as well as by two minima disposed at the frequencies
\begin{equation} \label{eq: min}
 \omega^{(min)}_{+,-}(k) = \pm \sqrt{\frac{-\mathcal{A}_1(k)-\sqrt{\mathcal{A}_1(k)^2-3\mathcal{A}_2(k)}}{3}}.
\end{equation}
\endnumparts
Note that Eqs. (43) account for the features of the dynamic
structure factor $S(k,\omega)$ at the frequencies $\omega^2 \ll 4
\Delta_4^2(k)$. Here, $\omega^{(max)}_{+,-}(k)$ are the positions
of the Brillouin symmetric components in $S(k,\omega)$. The dispersion relation $\omega_m(k)= c_s k$ with the adiabatic sound velocity $c_s$ can be recovered from the maxima of the longitudinal-current spectral density~(see, for instance, analysis of the experimental IXS data given in Ref.~\cite{Scopigno_RMP_2005}).
Nevertheless, since the longitudinal-current spectral density is proportional to $\omega^2 S(k,\omega)$, then it is obvious that relation~(\ref{eq: side})
with $\omega_c(k) \equiv \omega^{(max)}_{+,-}(k)$ will also
provide an information about the sound dispersion in a
liquid~\cite{Egelstaff1}. It is usually expected that both the
quantities $\omega_m(k)$ and $\omega_c(k)$ are close at
small values of $k$, and  the frequency $\omega_m(k)$ is extrapolated at
low-$k$ limit into the frequency of the Brillouin
doublet, i.e.
\begin{equation}\label{eq: equal}
  \lim_{k \to 0 } \sqrt{\frac{-\mathcal{A}_1(k)+\sqrt{\mathcal{A}_1(k)^2-3\mathcal{A}_2(k)}}{3}} =  c_s k.
\end{equation}
Further, the values of the frequencies
$\omega_m(k)$ and $\omega^{(max)}(k)$ diverge with increase of the wavenumber $k$~(see
discussion on p.~304 in Ref.~\cite{Boon/Yip_1991}).

Taking into account Eqs.~(\ref{eq: LT_M3}), (\ref{eq: LT_M2}) and (\ref{eq: cont_fraction}), we obtain the dispersion equation:
\begin{eqnarray} \label{eq: dispersion_equation}
    s^3 &+& \frac{2 \Delta_3^2(k) \sqrt{\Delta_4^2(k)}}{2\Delta_4^2(k) - \Delta_3^2(k)}s^2 \\
    &+& \left [ \Delta_1^2(k) + \Delta_2^2(k) + \frac{\Delta_2^2(k)\Delta_3^2(k)}{2\Delta_4^2(k)-\Delta_3^2(k)}  \right ] s + \frac{2\Delta_1^2(k)\sqrt{\Delta_4^2(k)}}{2\Delta_4^2(k) - \Delta_3^2(k)} = 0, \nonumber
\end{eqnarray}
which corresponds to the hydrodynamic dispersion equation~\cite{Mountain_1966}. Following the convergent Mountain's scheme for approximating solutions~\cite{Mountain_1966}, we find the correspondence between the hydrodynamic parameters and the frequency parameters for the low-$k$ limit:
\begin{equation} \label{eq: sound_vel}
    c_s k = \sqrt{ \Delta_1^2(k) + \Delta_2^2(k) + \frac{\Delta_2^2(k)\Delta_3^2(k)}{2\Delta_4^2(k)-\Delta_3^2(k)}},
\end{equation}
\begin{equation} \label{eq: D_T}
    D_T k^2 =  \frac{2 \Delta_1^2(k) \Delta_3^2(k) \sqrt{\Delta_4^2(k)}}{\Delta_1^2(k)[2\Delta_4^2(k)-\Delta_3^2(k)]+2\Delta_2^2(k)\Delta_4^2(k)}
\end{equation}
and
\begin{equation} \label{eq: Gamma}
    \Gamma k^2 = \frac{2\Delta_2^2(k)\Delta_3^2(k)\Delta_4^2(k) \sqrt{\Delta_4^2(k)}}{(2\Delta_4^2(k)-\Delta_3^2(k))[\Delta_1^2(k)(2\Delta_4^2(k)-\Delta_3^2(k))+2\Delta_2^2(k)\Delta_4^2(k)]}.
\end{equation}

Further, according to relation~(\ref{eq: Basic_a}), the intensity of the
central component of the dynamic structure factor $S(k,\omega)$
spectrum at the fixed wavenumber $k$ is
\begin{equation}\label{eq: Rayleigh_comp}
  S(k,\omega=0) = \frac{S(k)}{\pi} \frac{\Delta_2^2(k)\sqrt{\Delta_4^2(k)}}{\Delta_1^2(k)\Delta_3^2(k)}.
\end{equation}
In the hydrodynamic limit ($k \to 0$), this component transforms to the Rayleigh component:
\begin{equation}\label{eq: Rayleigh_hydr}
  S(k,\omega=0) = \frac{S(k)}{\pi}  \left( \frac{\gamma - 1}{\gamma} \right ) \frac{1}{D_T k^2},
\end{equation}
which is observed in the Brillouin light-scattering.

\section{Comparison with other theoretical schemes \label{sec: Correspondence}}

\subsection{Extended viscoelastic model \label{sec: ex_viscoel_model}}

It was revealed in Refs.~\cite{Scopigno_Li1,Scopigno_Li2} that the
three-peak lineshape of $S(k,\omega)$  of liquid metals  is well reproduced within the extended viscoelastic model (the so-called ``two relaxation time model'' according to terminology of
Ref.~\cite{Pilgrim_review,Aliotta_disp}; or ``two-time''
viscoelastic model -- in the notations of
Ref.~\cite{Bafile_collective}). In this model, the
scattering intensity is fitted by Eq.~(\ref{eq: DHOg}) for the
dynamic structure factor  with
the following approximation for the memory function:
\begin{eqnarray}\label{eq: M2_Sc} \label{eq: two_visc}
  \Delta_2^2(k) M_2(k,t) \simeq \sum_{j=D,\alpha,\mu}  b_{j}^2\; \mathrm{e}^{-t/\tau_j(k)}, \\
                        b_D^2(k) = (\gamma -1) \Delta_1^2(k),\ \ \ \tau_D(k) = (D_T k^2)^{-1}, \nonumber \\
                        b_{\alpha}^2 + b^2_{\mu}(k) =\Delta_L^2(k).\nonumber
\end{eqnarray}
Eq.~(\ref{eq: two_visc}) is an extension of the simple viscoelastic model~\cite{Copley_Lovesey}. Here,  the TCF $M_2(k,t)$  is approximated by the linear combination of the three
exponentially decaying functions responsible for
thermal fluctuations
\begin{equation}
m_{th}(k,t) \simeq \mathrm{e}^{-D_T k^2 t}
\end{equation}
 and for two viscous channels
 \begin{equation}
 m_{L}(k,t) \simeq \sum_{j=\alpha, \mu} b_{j}^2(k)\; \mathrm{e}^{-t/\tau_j(k)},
 \end{equation}
where  $b_j^2(k)$ and $\tau_j(k)$ are the strengths and the
relaxation times of the $\alpha$ and $\mu$ viscous regimes,
respectively~\cite{Scopigno_RMP_2005}. In the case of liquid
alkali metals, the specific-heat ratio $\gamma=c_P/c_V$ takes values
comparable with unity~\cite{Scopigno_RMP_2005}. As a result,
contribution of the first term (with the label $D_T$) in Eq.~(\ref{eq: two_visc}) to $M_2(k,t)$ is negligible. Therefore, the
time-dependence of the memory function $M_2(k,t)$ and the
lineshape of $S(k,\omega)$ will be determined mainly by the
parameters of two exponentially decaying functions associated with
the viscous modes~\cite{Pilgrim_review}. One needs to note that
there are no guidance and the microscopical relations to evaluate the characteristics of the viscous channels. Nevertheless, it was demonstrated~\cite{Scopigno_RMP_2005,Hosokawa1,Hosokawa2,Bove_Me} that  this extended viscoelastic model is capable to reproduce the experimental
$S(k,\omega)$-spectra within the wavenumber range $k \in [0.1;\;
0.5]\;k_m$ for various liquid metals (not only alkali
metals). Then the following question arises naturally:
\textit{Is there a correspondence between the theoretical model
presented in this work and the extended viscoelastic model?} A
simple way to verify a possible correspondence and then to answer
this question is to consider Eq.~(\ref{eq: LT_M2}) for the TCF
$\widetilde{M}_2(k,s)$ taking the quantity
\[
\xi(k)=\left |\frac{s^2}{4 \Delta_4^2(k)} \right |
\]
as a small parameter. By expanding the radicand in Eq.~(\ref{eq:
LT_M2}) over the small parameter $\xi$:
\begin{equation} \label{eq: expans_sqrt}
\sqrt{ 1 + \xi(k)} = 1 + \frac{\xi(k)}{2} - \frac{\xi^2(k)}{8} +
\ldots,
\end{equation}
one obtains directly from Eq.~(\ref{eq: LT_M3}) that
\begin{equation} \label{eq: M3_expan}
    \widetilde{M}_3(k,s) = -\frac{s}{2\Delta_4^2(k)} +
    \frac{1}{\sqrt{\Delta_4^2(k)}} + \frac{\xi(k)}{2\sqrt{\Delta_4^2(k)}}
    - \frac{\xi^2(k)}{8\sqrt{\Delta_4^2(k)}} + \ldots ,
\end{equation}
whereas Eq.~(\ref{eq: LT_M2}) takes the form:
\begin{equation}\label{eq: expansion_M2}
  \widetilde{M}_2(k,s) \simeq \sum_{j} \frac{a_j^2(k)}{s + \tau_j^{-1}(k)},
\end{equation}
or, equivalently,
\begin{eqnarray}\label{eq: M2_t_expans}
  M_2(k,t) &\simeq& \sum_{j} a_j^2(k)\; \mathrm{e}^{-t/\tau_j(k)}, \\
   & & \sum_j a_j^2(k) = 1, \nonumber \\
   & &j=1,\;2,\;3,\;5,\;\ldots. \nonumber
\end{eqnarray}
Here, the weight coefficients $a_j^2(k)$ and the relaxation times
$\tau_j(k)$ are defined through the frequency parameters
$\Delta_3^2(k)$ and $\Delta_4^2(k)$. Thus, we obtained in the framework of our theoretical model an approximate expansion of the function $M_2(k,t)$ by the exponential contributions. The amount of the Lorentzian
functions in (\ref{eq: expansion_M2}) and/or the amount of the
exponential functions in (\ref{eq: M2_t_expans}) is determined by
the amount of terms taken into account in expansion~(\ref{eq: expans_sqrt}).

For example, assuming that $\xi(k) \to 0$, one
obtains
\begin{equation}
\widetilde{M}_3(k,s) \simeq \frac{2\sqrt{\Delta_4^2(k)} -
s}{2\Delta_4^2(k)}
\end{equation}
and
\begin{equation} \label{eq: ve}
\widetilde{M}_2(k,s) \simeq \frac{1}{s+\tau^{-1}(k)}, \ \ \ M_2(k,t) \simeq \mathrm{e}^{- t/\tau(k)}
\end{equation}
with the relaxation time
\begin{equation} \label{eq: ve_times}
\tau^{-1}(k) \simeq
\frac{2\sqrt{\Delta_4^2(k)}\Delta_3^2(k)}{2\Delta_4^2(k)-\Delta_3^2(k)}.
\end{equation}
Eq.~(\ref{eq: ve}) is equivalent to Eq.~(\ref{eq: viscoel_M2}) and corresponds to the \textit{simple viscoelastic model}~\cite{Copley_Lovesey}.

Further, expansion~(\ref{eq: M2_t_expans}) at $j=3$ is similar to
the \textit{extended viscoelastic model} with the memory function
$M_2(k,t)$ represented by Eq.~(\ref{eq: two_visc}). The inverse
relaxation times for this case can be estimated as follows
\begin{equation} \label{eq: rel_time_visco}
  \tau_{\alpha,\mu}^{-1}(k) \simeq \sqrt{\Delta_4^2(k)} \pm \sqrt{\Delta_4^2(k) - \Delta_3^2(k)},
\end{equation}
whereas the weight coefficient (say, for the $\alpha$th contribution) is
\begin{equation} \label{eq: weight}
  b_{\alpha}^2(k) \simeq \frac{1}{2} + \frac{1}{2} \left (\sqrt{1 -\Delta_3^2(k)/\Delta_4^2(k)} \right )^{-1}.
\end{equation}

Thus, we have identified relationship between the presented theory and the viscoelastic models.
Although both the simple and extended viscoelastic models do not satisfy the high-order sum rules, these models can yield the correct results for the spectral features at the frequencies much smaller than $2\sqrt{\Delta_4^2(k)}$. Therefore, the extended viscoelastic model can be used as a sufficiently convenient approximation to study the relaxation process with the characteristic time scales larger than $1/(2\sqrt{\Delta_4^2(k)})$.
This is directly evident from obtained above the simple analytical expansion for the TCF $M_2(k,t)$. Further, Eqs. (\ref{eq: ve_times}), (\ref{eq: rel_time_visco}) and (\ref{eq: weight}) set correspondence between the parameters of the viscoelastic models and the frequency parameters [the spectral moments of $S(k,\omega)$].

\subsection{Hydrodynamic limit \label{sec: hydrodynamics}}

Relation for the dynamic structure factor $S(k,\omega)$ at the
hydrodynamic limit [see Eq.~(\ref{eq: hydrodynamics})] can be
exactly rewritten as the ratio of the biquadratic polynomial to
the bicubic polynomial: \numparts \label{eq: hydro_polin}
\begin{equation}\label{eq: hydro_ratio}
  2\pi \frac{S(k,\omega)}{S(k)} = \frac{a_1(k) \omega^4 + a_2(k) \omega^2 +a_3(k)}{\omega^6 + b_1(k) \omega^4 + b_2(k) \omega^2 + b_3(k)},
\end{equation}
where the coefficients of the polynomials are defined as follows
\begin{equation}\label{eq: a1}
  a_1(k) = \frac{2}{\gamma} \left [ (\gamma - 1) D_T k^2 + \Gamma k^2    \right ],
\end{equation}
\begin{eqnarray}\label{eq: a2}
  a_2(k) &=& \frac{2}{\gamma} \left \{ 2 (\gamma - 1) D_T k^2 \left [ (\Gamma k^2)^2 - (c_s k)^2 \right ] \right . \\
   & & \left .  + (\Gamma k^2) \left [ (\Gamma k^2)^2 + (c_s k)^2 + (D_T k^2)^2 \right ]    \right \},\nonumber
\end{eqnarray}
\begin{eqnarray}\label{eq: a3}
  a_3(k) &=& \frac{2}{\gamma} D_T k^2 \left [ (\Gamma k^2)^2 + (c_s k)^2 \right ] \\
  &\times& \left \{ (\gamma - 1) \left [ (\Gamma k^2)^2 + (c_s k)^2 \right ]         + (D_T k^2)(\Gamma k^2)    \right \},\nonumber
\end{eqnarray}
\begin{equation}\label{eq: b1}
  b_1(k) = 2 (\Gamma k^2)^2 - 2 (c_s k)^2 + (D_T k^2)^2,
\end{equation}
\begin{eqnarray}\label{eq: b2}
  b_2(k) &=& 2 (D_T k^2)^2 \left [ (\Gamma k^2)^2 - (c_s k)^2\right ]\\
  & & + \left [ (\Gamma k^2)^2 + (c_s k)^2 \right ] ^2,\nonumber
\end{eqnarray}
\begin{equation}\label{eq: b3}
  b_3(k) =   (D_T k^2)^2 \left [ (\Gamma k^2)^2 + (c_s k)^2 \right ]^2.
\end{equation}
\endnumparts
As seen from Eqs.~(\ref{eq: hydro_ratio}), the coefficients in Eq.~(\ref{eq: hydro_ratio}) at the low-$k$ limit have the following $k$-dependence: $a_1(k) \sim k^2$, $a_2(k) \sim k^4$, $a_3(k)
\sim k^6$, $b_1(k) \sim k^2$, $b_2(k) \sim k^4$ and $b_3(k) \sim
k^6$.

On the other hand, inserting expansion~(\ref{eq: expans_sqrt})
into relation~(\ref{eq: Basic_a}) one obtains
\numparts \label{eq: Sk_om}
\begin{equation}
 \label{eq: Basic_expan}
 2\pi\frac{S(k, \omega)}{S(k)} = \frac{\mathcal{B}_1(k) \omega^4 + \mathcal{B}_2(k) \omega^2 + \mathcal{B}_3(k)}
 {\omega^6 +\mathcal{A}_1(k) \omega^4 +\mathcal{A}_2(k) \omega^2 +\mathcal{A}_3(k)},
\end{equation}
where
\begin{equation}\label{eq: B1}
  \mathcal{B}_1(k) = \frac{\Delta_1^2(k) \Delta_2^2(k) \Delta_3^2(k)}{16 \Delta_4^2(k) \sqrt{\Delta_4^2(k)} (\Delta_4^2(k) - \Delta_3^2(k) )},
\end{equation}
\begin{equation}\label{eq: B2}
  \mathcal{B}_2(k) = \frac{\Delta_1^2(k) \Delta_2^2(k) \Delta_3^2(k)}{4 \sqrt{\Delta_4^2(k)} (\Delta_3^2(k) - \Delta_4^2(k) )},
\end{equation}
\begin{equation}\label{eq: B3}
  \mathcal{B}_3(k) = \frac{2 \Delta_1^2(k) \Delta_2^2(k) \Delta_3^2(k) \sqrt{\Delta_4^2(k)}}{\Delta_4^2(k) - \Delta_3^2(k) },
\end{equation}
\endnumparts
and the coefficients $\mathcal{A}_1(k)$, $\mathcal{A}_2(k)$
and $\mathcal{A}_3(k)$ are defined by Eqs.~(\ref{eq: coeff_A}).

Thus, theoretical model~(\ref{eq: Basic_a}) for the dynamic structure factor can be reduced to the hydrodynamic result at long wavelengths and low frequencies.

\subsection{Generalized hydrodynamics \label{sec: gen_hydro}}

The basic idea of the generalized hydrodynamics consists in
introducing the $k$-dependent relaxation processes into the
linearized hydrodynamics equations~\cite{Boon/Yip_1991}. Then, the
dynamic structure factor $S(k,\omega)$ will be modified, and it
takes a more complicated form in comparison with hydrodynamic
equation for $S(k,\omega)$, i.e. Eq.~(\ref{eq: hydrodynamics}). As
shown in Refs.~\cite{Boon/Yip_1991,Aliotta_disp}, the dispersion
relation for this case can be written as follows \numparts
\label{eq: GH}
  \begin{equation}\label{eq: 1}
    \omega(k) = c_s k = c_{s,0} k \sqrt{ S + \sqrt{S^2 + \frac{1}{(c_{s,0} k \tau)^2}}},
  \end{equation}
  \begin{equation}\label{eq: 2}
    S = \frac{1}{2} \left [ \frac{c_{s,\infty}^2}{c_{s,0}^2}   - \frac{1}{(c_{s,0} k \tau)^2} \right ].
  \end{equation}
\endnumparts
Here, $c_{s,0}$ is the low frequency speed of sound, $c_s =
\omega/k$ is the speed of sound at the given frequency $\omega$,
$c_{s,\infty}$ is the high frequency sound velocity, and $\tau$ is
the relaxation time~\footnote{Do not confuse the quantity $S$
in Eq.~(\ref{eq: 1}) with the static structure factor
$S(k)$.}.

As seen, dispersion relation~(\ref{eq: 1}) is similar to
dispersion law~(\ref{eq: side}) with only difference of the sign
under the inner radicant. Note that the negative contribution
under the inner radicant  in Eq.~(\ref{eq: side}), i.e.
$-3\mathcal{A}_2(k)$, provides an impact into the decreasing
$\omega_s(k)$ at the wavenumbers from the range  $k \in (k_m/2;\;k_m)$. Further, the
quantity $S$ in Eq.~(\ref{eq: 1}) is identified with the
coefficient $\mathcal{A}_1(k)$ in Eq.~(\ref{eq: side}):
\[
S \to \mathcal{A}_1(k). \] Note that the coefficient
$\mathcal{A}_1(k)$ takes the negative values (see, for example, Tab.~\ref{tab:
Ak} with values of this coefficient for liquid lithium).

\subsection{High-$k$ limit \label{sec: free_part_limit}}

The regime of the free-particle dynamics arises at the large
wavenumbers, i.e. $k > k_m$, corresponding to wavelengths
comparable with and smaller than the mean free path of a moving
particle. Time- and spatial-scales are so short that there is, in
fact, no any vibrational dynamics. The dynamic structure
factor for the regime takes a sense of the self-dynamic structure
factor $S_{s}(k,\omega)$ measurable in INS and characterizing a
single-particle dynamics~\cite{Egelstaff1}. The static structure
factor for this $k$-range approaches unity, i.e. $S(k) \to
1$~\cite{Copley_Lovesey}. As a result, all the contributions in
expressions for the frequency parameters, $\Delta_2^2(k)$,
$\Delta_3^2(k)$ and etc., responsible for the particles
interaction can be omitted. Then, taking into account Eqs.~(28) one obtains
\begin{equation}\label{eq: free_dyn_freq}
  \Delta_1^2(k) = (v_T k)^2, \  \Delta_2^2(k) = 2 \Delta_1^2(k), \ \Delta_3^2(k) = 3 \Delta_1^2(k), \  \ldots.
\end{equation}
Substituting the frequency parameters into fraction~(\ref{eq: cont_fraction}) one has
\begin{equation} \label{eq: cont_fraction_free}
\widetilde{F}(k,s)=
\frac{1}{\displaystyle s+\frac{\Delta_{1}^{2}(k)}{\displaystyle
s+\frac{2\Delta_{1}^{2}(k)}{\displaystyle
s+\frac{3\Delta_{1}^{2}(k)}{\displaystyle s+\ddots}}}}.
\end{equation}
This is continued fraction representation of the Laplace
transform of the Gaussian function
\begin{equation}\label{eq: Gaussian_a}
  F(k,t) = \exp \left ( - \frac{ (v_T k)^2 t^2}{2}  \right ).
\end{equation}
Then, taking into account Eq.~(\ref{eq: dsf}), one obtains the correct
result for the dynamic structure factor at the high-$k$ limit:
\begin{equation} \label{eq: Gaussian1}
  S(k,\omega) = \sqrt{\frac{1}{2\pi}} \frac{1}{v_T k} \exp\left( - \frac{\omega^2}{2 (v_T k)^2}  \right ), \nonumber
\end{equation}
which was mentioned in Sec.~\ref{sec: intr}.

We see that expression (\ref{eq: Gaussian1}) is the rigorous theoretical result obtained in a self-consistent manner from the exact relation (\ref{eq: free_dyn_freq}) for the frequency parameters. Therefore, relation (\ref{eq: Basic_a}) for the dynamic structure factor transforms into relation (\ref{eq: Gaussian1}), when condition (\ref{eq: time_scales_equality}) for the frequency parameters changes to (\ref{eq: free_dyn_freq}).

\section{Dynamic structure factor of liquid lithium near melting \label{sec: lithium}}

The first studies of the microscopic collective ion dynamics of liquid lithium by means of inelastic scattering methods were performed by  De Jong and Verkerk~\cite{Verkerk_INS}, Burkel and Sinn~\cite{Burkel_IXS}. Then, IXS studies of
this liquid metal were carried out by Scopigno \textit{et al.}  for the more extended range of the wavenumbers.

\begin{figure}[ht]
\centering
\includegraphics[width=0.7\linewidth]{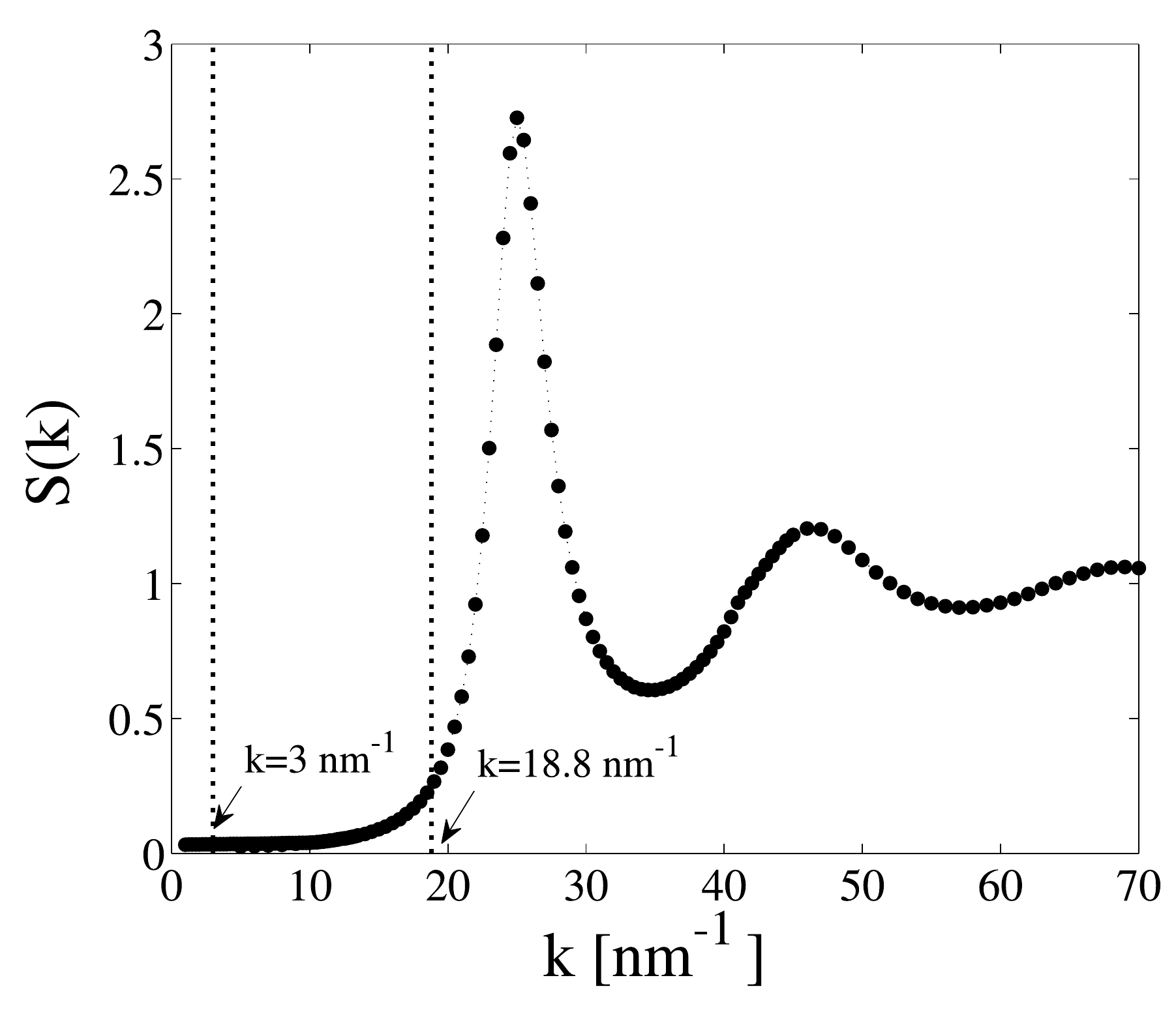}
\caption{Static structure factor of liquid lithium at the
temperature $T=463$~K measured by X-ray diffraction; data taken
from Ref.~\cite{Waseda}. Two vertical dotted lines mark out the
wavenumber range considered in this study. The main peak of the
static structure factor $S(k)$ is located at $k_m \simeq
24.4$~nm\;$^{-1}$.} \label{fig: Sk}
\end{figure}

The features of the density fluctuations dynamics in a liquid depend
on a spatial scale, where these fluctuations emerge. Therefore, it is
convenient to measure the spatial scale in terms of the location
of the first peak $k_m$ in the static structure factor $S(k)$.  In
the given study, the equilibrium density fluctuations dynamics in
liquid lithium at the temperature $T=475$~K is considered. For liquid lithium at
this temperature, the quantity $k_m$ equals to $\simeq
24.4$~nm$^{-1}$ (see Fig.~\ref{fig: Sk}). Thus, the extended
wavenumber range $k \in [3.0;\;18.8]$~nm$^{-1}$  is covered by our
study. According to the classification given in
Ref.~\cite{Scopigno_RMP_2005} (see on p.~927-930), the wavenumber $k=3.0$~nm$^{-1}$ corresponds to $k/k_m=0.12$ and defines the
lower boundary of a hypothetical isothermal dynamical region, whereas the wavenumber $k=18.8$~nm$^{-1}$ is identical to $k/k_m=0.77$ and is associated with the upper boundary of the so-called generalized hydrodynamic region~\cite{Scopigno_RMP_2005}. Although the theory presented in Sec.~\ref{sec: formalism} should be also valid for longer wavelengths, $k=3.0$~nm$^{-1}$ is the smallest wavenumber with  available IXS data for liquid lithium~\cite{Scopigno_Li1}. On the other hand, the wavenumber $k=18.8$~nm$^{-1}$ corresponds to the shortest wavelengths at which the collective acoustic-like dynamics is still supported in this liquid and expression (\ref{eq: Basic_a}) for the dynamic structure factor $S(k,\omega)$ is valid.

To compare the theoretical classical dynamic structure factor
$S(k,\omega)$ with the scattering intensity $I(k,\omega)$, one needs to
account for the quantum effects by means of the detailed balance
condition
\numparts
\label{eq: comparison}
\begin{equation}\label{eq: detailed_balance}
  S_q(k,\omega) = \frac{\hbar \omega/k_B T}{1-\exp(-\hbar \omega /k_B T)} S(k,\omega),
\end{equation}
where $S_q(k,\omega)$ is the quantum-mechanical structure factor. Then, the experimental resolution effects must be included into consideration through the convolution
\begin{equation}\label{eq: concolution}
  I(k,\omega) = E(k)\int R(k,\omega - \omega ') S_q(k,\omega') d \omega',
\end{equation}
\endnumparts
where $E(k)$ and $R(k,\omega)$ are the characteristics of the IXS
experiment. The quantity $E(k)$ depends on analyzer efficiency and
on the atomic form factor~\cite{Scopigno_RMP_2005}, whereas
$R(k,\omega)$ is the experimental resolution function.
\begin{figure}[ht]
\centering
\includegraphics[width=0.7\linewidth]{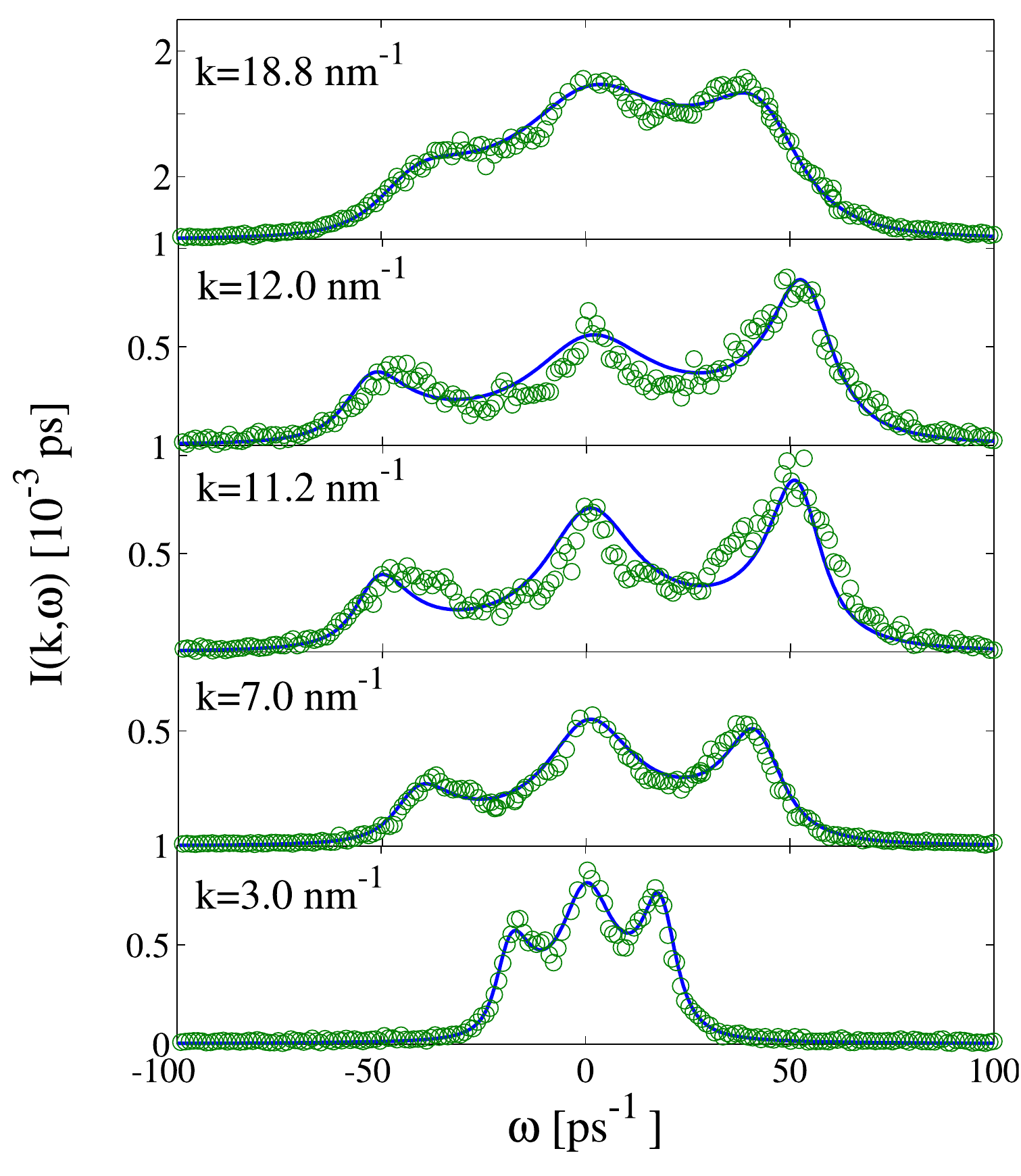}
\caption{Scattering intensity $I(k,\omega)$ of liquid lithium at the
temperature $T=475$~K and several values of the wavenumber
$k<k_m$, where $k_m=24.4$~nm$^{-1}$ corresponds to the first peak
position of the static structure factor $S(k)$. Open circles
($\circ$) represent experimental IXS
data~\cite{Scopigno_RMP_2005}, full lines ($-$) depict the
theoretical dynamical structure factor $S(k,\omega)$ [see
Eq.~(\ref{eq: Basic_a})] modified to takes into account
the quantum-mechanical detailed balance condition and the
experimental resolution effects.}\label{fig: dsf}
\end{figure}
\begin{table}[h]
\caption{\label{tab: Ak} Values of the frequency parameters $\Delta_1^{2}(k)$, $\Delta_2^{2}(k)$, $\Delta_3^{2}(k)$, $\Delta_4^{2}(k)$ (in units of $\times
10^{24}$~s$^{-2}$) and values of the coefficients $\mathcal{A}_1(k)$ ($\times
10^{26}$~s$^{-2}$), $\mathcal{A}_2(k)$ ($\times 10^{52}$~s$^{-4}$), $\mathcal{A}_3(k)$ ($\times 10^{78}$~s$^{-6}$) of
the polynomial in relation~(\ref{eq: Basic_a}) for the dynamic
structure factor~$S(k,\omega)$. }
\begin{indented}
\item[]\begin{tabular}{@{}ccccccccc} \br
    $k$  (nm$^{-1}$) & $\Delta_1^{2}(k)$  & $\Delta_2^{2}(k)$  & $\Delta_3^{2}(k)$ & $\Delta_4^{2}(k)$ & $\mathcal{A}_1(k)$   & $\mathcal{A}_2(k)$  & $\mathcal{A}_3(k)$   \\
    \mr
    $3.0$  & $183\pm10$  & $210\pm30$  & $1500\pm230$  & $17731\pm4500$ & $-6.61$  & $11.0$  & $4.5$    \\
    $7.0$ & $845\pm30$  & $1109\pm150$  & $7110\pm1350$  & $87905\pm23500$ & $-33.78$ & $294.8$ & $446.2$  \\
    $11.2$ & $1346\pm50$  & $1500\pm220$  & $10000\pm3100$  & $179700\pm43000$ & $-52.07$ & $681.0$ & $1071.8$ \\
    $12.0$ & $1545\pm80$  & $1609\pm240$  & $12000\pm3000$  & $157600\pm40500$ & $-54.48$ & $729.8$ & $2369.9$ \\
    $18.8$ & $971\pm60$  & $1589\pm210$  & $11100\pm2300$  & $68540\pm20000$ & $-32.77$ & $316.4$ & $2027.0$ \\
\br
\end{tabular}
\end{indented}
\end{table}
The dynamic structure factor $S(k,\omega)$ and the scattering intensity
$I(k,\omega)$ of liquid lithium at $T=475$~K were defined by means
of Eqs.~(\ref{eq: Basic_a}) and Eqs.~(\ref{eq: detailed_balance}),
(\ref{eq: concolution}) at the wavenumbers $k=3.0$, $7.0$,
$11.2$, $12.0$ and $18.8$~nm$^{-1}$. To determine $S(k,\omega)$
by means of relation~(\ref{eq: Basic_a}) we first need to compute the four
frequency parameters $\Delta_1^2(k)$, $\Delta_2^2(k)$, $\Delta_3^2(k)$ and $\Delta_4^2(k)$. This can be done by two independent methods.

Let us consider the \textit{first method} used by us to determine these parameters for the case of liquid lithium at the considered thermodynamic state. The frequency parameter $\Delta_1^2(k)$ has been computed exactly by means of Eq.~(\ref{eq: D1}) [the first equality in this equation] with the experimental values of the static structure factor $S(k)$ from Ref.~\cite{Waseda}.
The values of the parameter $\Delta_2^2(k)$  were
determined from the first equality of Eq.~(\ref{eq: D2}) with the pseudopotential $u(\mathbf{r})$ proposed by Gonzalez \textit{et al.}~\cite{Gonz_2001} and with the radial distribution function $g(r)$ computed on the basis of this potential.
It is important to note that the theoretical radial distribution function is consistent with the X-ray diffraction data~\cite{Waseda}. Further, although the frequency parameters $\Delta_3^2(k)$ and
$\Delta_4^2(k)$ can be also computed theoretically through the
integral expressions  with the three- and four-particle
distribution functions [see, for example, Eq.~(\ref{eq: D3})], this is difficult to implement. Therefore, one can take these two parameters
as the fitting parameters. The first way is to calculate the frequency
moments $\omega^{(6)}(k)$ and  $\omega^{(8)}(k)$ for the
\textit{experimental} spectra of the dynamic structure factor $S(k,\omega)$ [see Eq.~(\ref{eq: freq_moments})],
and then to find the values of $\Delta_3^2(k)$ and $\Delta_4^2(k)$ by means of
sum rules~(\ref{eq: Delta3}) and (\ref{eq:
Delta4}).~\footnote{According to the method, these moments are
evaluated for the dynamic structure factor, which has to be
extracted by deconvolution of the experimental scattering
intensity and the experimental resolution function. For the case
of liquid metals, the dynamic structure factor at finite
wavenumbers decays (in frequency) to zero over $100$~THz range.
Therefore, all the moments have to be finite and definable by
this method. Computational problems can be here only due to
deconvolution procedure. Nevertheless, one needs
to note that the resolution function is well approximated by the Dirac
delta-function for the case of INS data. Therefore, the scattering  intensity is proportional to the dynamic scattering function, and the method for evaluating the
spectral moments is expected to be efficient.} We have evaluated these parameters, $\Delta_3^2(k)$ and $\Delta_4^2(k)$, by a different way.
Namely,  if one
associates the parameter $\Delta_4^2(k)$ with the correct
(experimental) value of $S_{exp}(k,\omega)$ at $\omega=0$ in accordance with
relation~(\ref{eq: Basic_a}), i.e.
\begin{equation}
\Delta_4^2(k) = \left [  \frac{\pi S_{exp}(k,\omega=0)}{S(k)} \frac{\Delta_1^2(k) \Delta_3^2(k)}{\Delta_2^2(k)}  \right ]^2,\label{eq_delta4}
\end{equation}
then the quantity $\Delta_3^2(k)$ must be taken directly as fitting parameter. Thus, the two parameters, $\Delta_3^2(k)$ and $\Delta_4^2(k)$,
are used as adjustable in such the realization of the theoretical model~(\ref{eq: Basic_a}) for $S(k,\omega)$. To our knowledge, there is no other
theory for the dynamic structure factor of liquids (even of simple
liquids), which operates without fitting parameters, or applies one or two fitting parameters only and yields, herewith,
satisfactory agreement with experimental IXS/INS data for the
wavenumber range. Further, we recall that the frequency parameters
are not the special model parameters, but are the quantities,
that determine the sum-rules of the scattering law. Consequently, if the
theoretical scheme is capable to reproduce all the features of the
scattering law and the scheme satisfies the sum-rules, then
theoretical and experimental values of the frequency parameters
must be comparable or identical.

The first six frequency parameters, $\Delta_1^2(k)$,
$\Delta_2^2(k)$, $\Delta_3^2(k)$, $\Delta_4^2(k)$, $\Delta_5^2(k)$
and $\Delta_6^2(k)$, were also evaluated by means of the \textit{second method}.
In accordance with this method, the configuration data obtained from the simulations with the given interparticle potential $u(\mathbf{r})$ is required only. Then, the frequency moments $\omega^{(j)}(k)$ and the frequency parameters $\Delta_j^2(k)$ can be determined by means of the original definitions: Eq.~(\ref{eq: frequency_parameter}) with relations
(\ref{eq: density_fluct}) and (\ref{eq: variables}). The computational details are given in Appendix.  This method yields results for the low-order frequency moments and parameters with high accuracy, that is confirmed by agreement of the results with the data for the experimental static structure factor $S(k)$ and with the exact theoretical values for $\Delta_1^2(k)$ and $\Delta_2^2(k)$ evaluated from Eqs.~(\ref{eq: D1}) and (\ref{eq: D2}), respectively. Because of the numerical differentiation procedure used in this method, the magnitude of the statistical error of an evaluated frequency parameter $\Delta_j^2(k)$ increases with increasing order $j$, where $j =3,\;4,\;5,\;\ldots$. Nevertheless, we find that the values of $\Delta_3^2(k)$ and $\Delta_4^2(k)$ agree within the confidence interval with the results obtained  from the fitting procedure in accordance with the first method.
This confirms that the values of the frequency parameters $\Delta_1^2(k)$,
$\Delta_2^2(k)$, $\Delta_3^2(k)$ and $\Delta_4^2(k)$ obtained by the first method can be adopted; the numerical values of these parameters are given in Tab.~\ref{tab: Ak}.  The results for the frequency parameters from both the methods will be presented in Fig.~\ref{fig: Deltas}.

In Fig.~\ref{fig: dsf}, the theoretical scattering intensity
$I(k,\omega)$ is compared with the experimental IXS
data~\cite{Scopigno_RMP_2005}. As seen, the presented theory
with expression~(\ref{eq: Basic_a}) for the dynamic
structure factor has excellent agreement with experimental data.
The theoretical intensity $I(k,\omega)$ accounts for the
spectral features. The values of the coefficients
$\mathcal{A}_1(k)$, $\mathcal{A}_2(k)$ and $\mathcal{A}_3(k)$ are given in Tab.~\ref{tab: Ak}.
\begin{figure}[ht]
\centering
\includegraphics[width=1.0\linewidth]{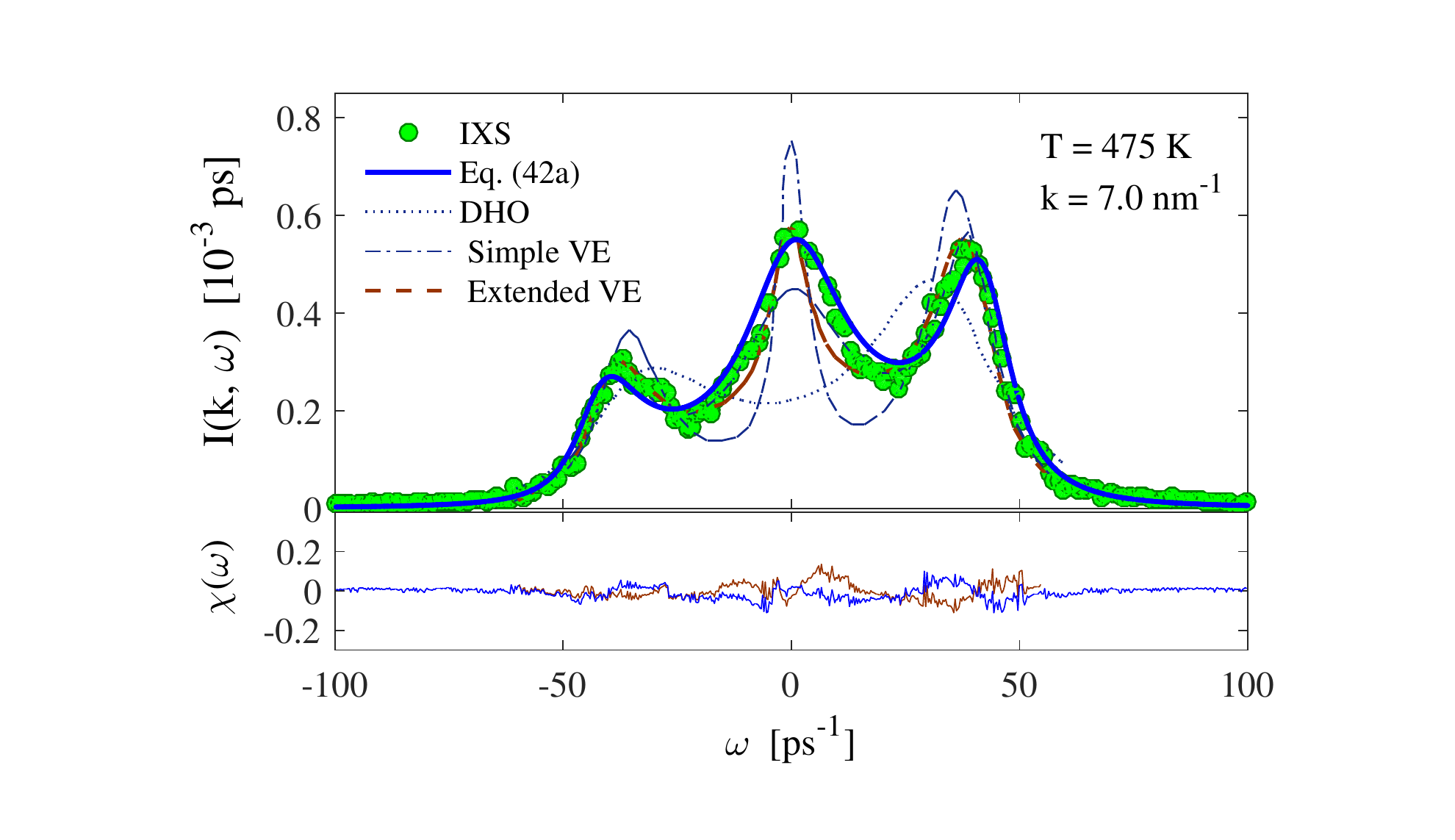}
\caption{Top: Scattering intensity of liquid lithium at $T=475$~K and $k=7.0$~nm$^{-1}$: IXS data (full circles) and results for the DHO model (dotted line), the simple viscoelastic model ($- \cdot -$), the extended viscoelastic model (dashed line) and the results of the theoretical model~(\ref{eq: Basic_a}). IXS data and results for DHO, simple and extended viscoelastic models are taken from Ref.~{\cite{Scopigno_Li1}}.  Bottom: Difference between the experimental data and the theoretical values, $\chi(\omega) = I_{IXS}(k,\omega) - I_{th}(k,\omega)$, computed for the extended viscoelastic model and for the theoretical model ~(\ref{eq: Basic_a}).}\label{fig: comp}
\end{figure}
\begin{figure}[ht]
\centering
\includegraphics[width=0.7\linewidth]{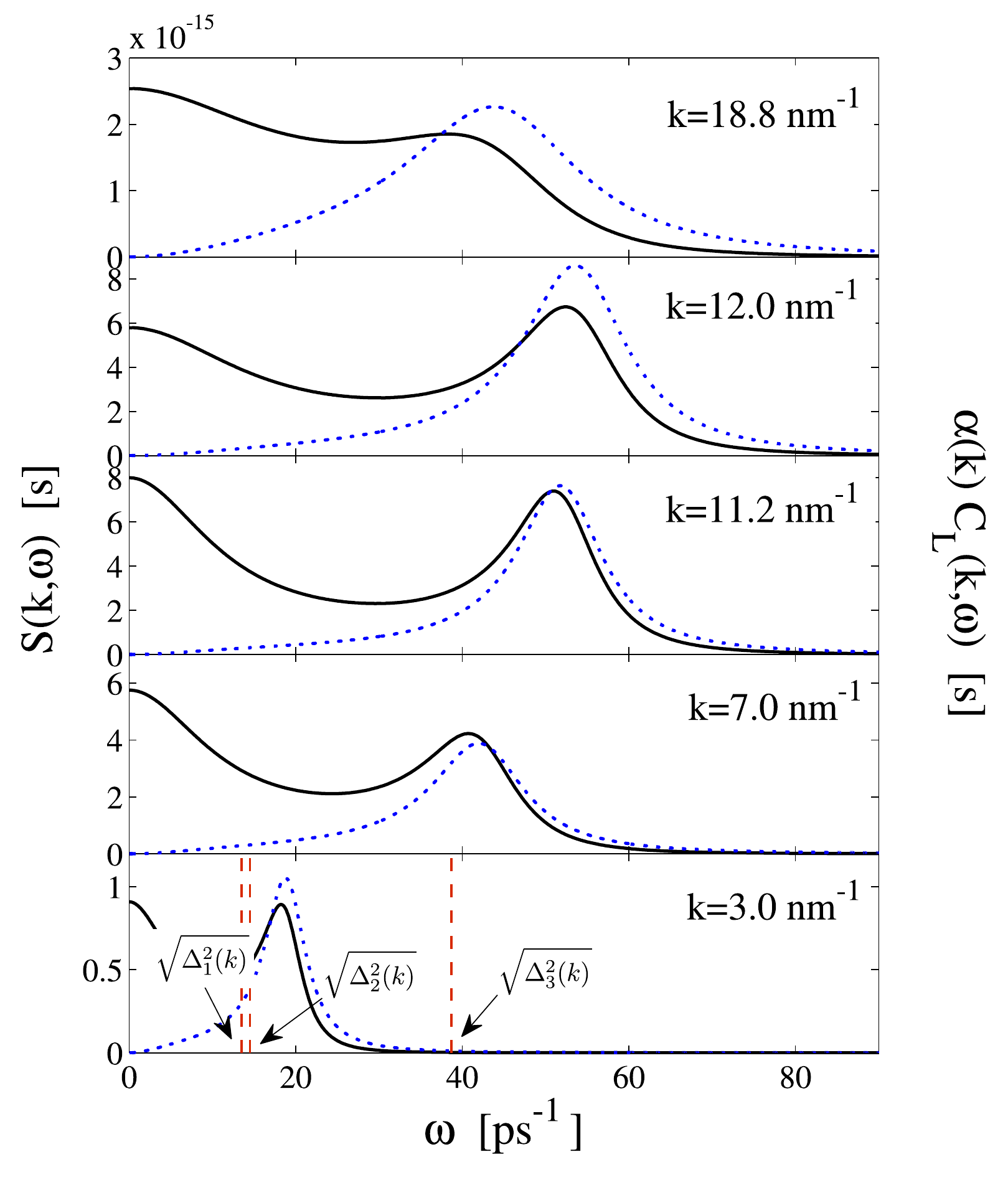}
\caption{Dynamic structure factor $S(k,\omega)$ (solid line) and
 scaled longitudinal current correlation function
$\alpha(k)C_L(k,\omega)$ (dotted line) of liquid lithium at the
same temperature and fixed wavenumbers as in Fig.~\ref{fig: dsf}.
The dimensionless coefficient $\alpha(k)$ takes the values $5
\cdot 10^{-5}$, $5 \cdot 10^{-5}$, $15 \cdot 10^{-5}$, $2 \cdot
10^{-4}$, and $3 \cdot 10^{-3}$ at $k=3.0$, $7.0$, $11.2$, $12.0$
and $18.8$~nm$^{-1}$, respectively. Red vertical dashed lines for
the bottom plot correspond to the frequencies
$\sqrt{\Delta_1^2(k)}$, $\sqrt{\Delta_2^2(k)}$ and
$\sqrt{\Delta_3^2(k)}$ at $k=3.0$~nm$^{-1}$.}\label{fig: dsf_curr}
\end{figure}

We note that efficiency of some theoretical models to analyze the same IXS data was verified previously by Scopigno \emph{et al.} in Ref.~\cite{Scopigno_Li1}. Comparison of experimental data and theoretical results was carried out for the specific spectrum at $k=7.0$~nm$^{-1}$. As was clearly demonstrated (Fig. 5 in Ref.~\cite{Scopigno_Li1}), both the damped harmonic oscillator model and the simple viscoelastic model cannot reproduce the lineshape of the scattering intensity.  The agreement with the experimental data appears only for the extended viscoelastic model, where the time dependence of the TCF $M_2(k,t)$ is approximated by the linear combination of
the three exponential decay laws associated with thermal
and viscous fluctuations. In Fig.~\ref{fig: comp}, we reproduce all the data from Fig. 5 of Ref.~\cite{Scopigno_Li1} supplemented by our theoretical results. Although the extended viscoelastic model of Ref.~\cite{Scopigno_Li1} and our model with Eq. (\ref{eq: Basic_a}) do not yield the identical lineshapes of $I(k,\omega)$, both the models are characterized by a comparable agreement with the experimental data.
Further,  as it was demonstrated in Sec.~\ref{sec: Correspondence}, our theory
does not contradict to the extended viscoelastic model and  can provide justification for the expansion of the TCF $M_2(k,t)$ over exponential
functions. For the case of liquid lithium with the estimated frequency parameters $\Delta_3^2(k)$ and $\Delta_4^2(k)$ at the considered wavenumbers,  we find from Eq.~(\ref{eq: rel_time_visco}) that the first relaxation time
$\tau_{\alpha}(k)$
takes the values from the range $[0.1; 1]$~ps,
whereas the second relaxation process with $\tau_{\mu}(k)$ occurs over time
scales $\sim 0.01$~ps. These estimates are in agreement with
results obtained in Refs.~\cite{Scopigno_Li1,Scopigno_Li2}
with the extended viscoelastic model.

\begin{figure}[ht]
\centering
\includegraphics[width=0.7\linewidth]{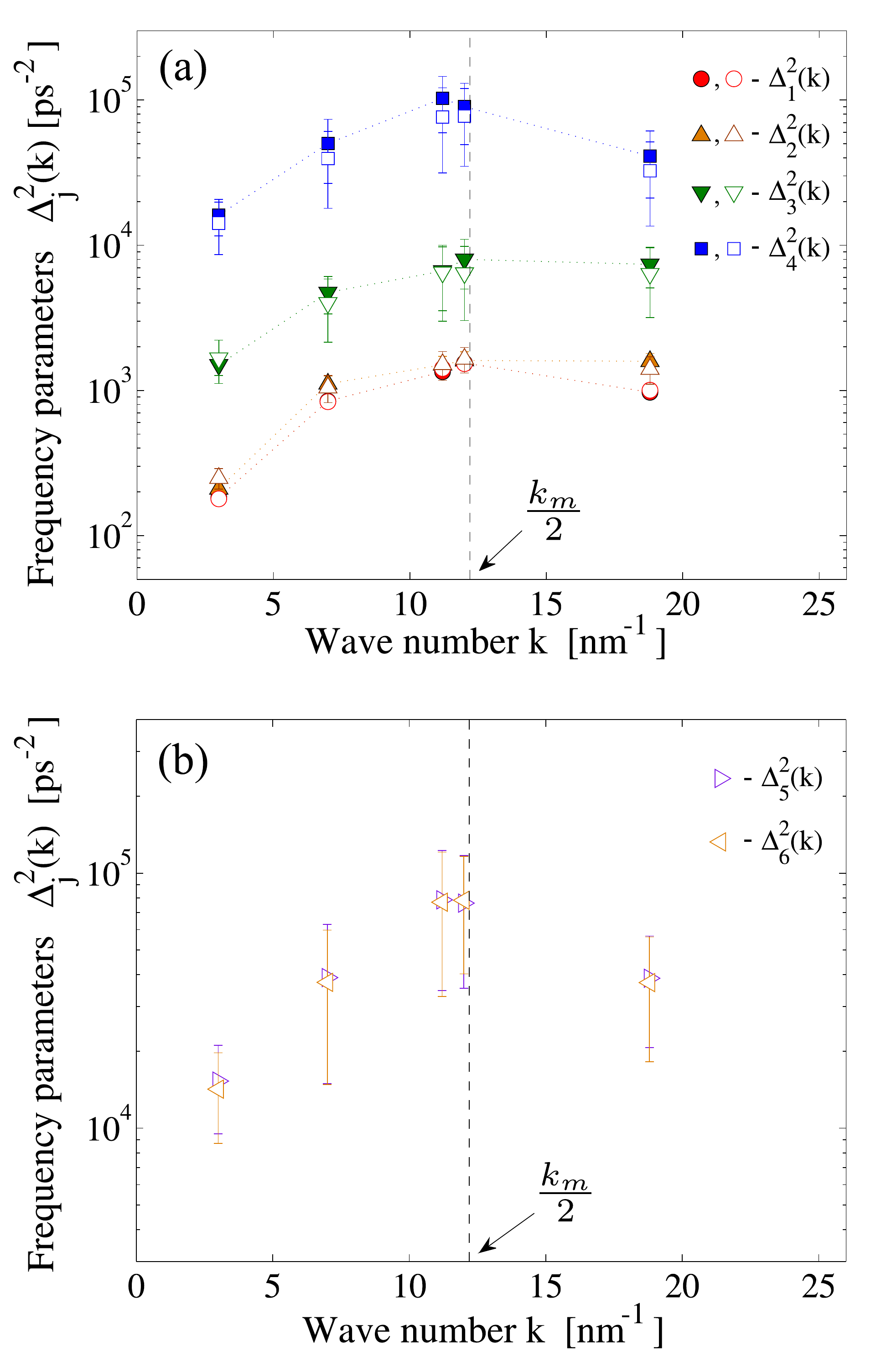}
\caption{Wavenumber dependence of the frequency parameters
$\Delta_1^2(k)$, $\Delta_2^2(k)$, $\Delta_3^2(k)$,
$\Delta_4^2(k)$, $\Delta_5^2(k)$ and $\Delta_6^2(k)$ in
logarithmic scale. Dashed vertical line corresponds to the
wavenumber $k=k_m/2\simeq 12.2$~nm\;$^{-1}$. Full markers
correspond to the theoretical values evaluated by the first way
(see the text before Eq.~(\ref{eq_delta4})), whereas transparent
markers indicate the values estimated theoretically by the second
way -- on the basis of Eqs.~(\ref{eq: variables}) and (\ref{eq:
frequency_parameter}) with the configuration data (for details,
see Appendix).}\label{fig: Deltas}
\end{figure}
\begin{figure}[ht]
\centering
\includegraphics[width=1.2\linewidth]{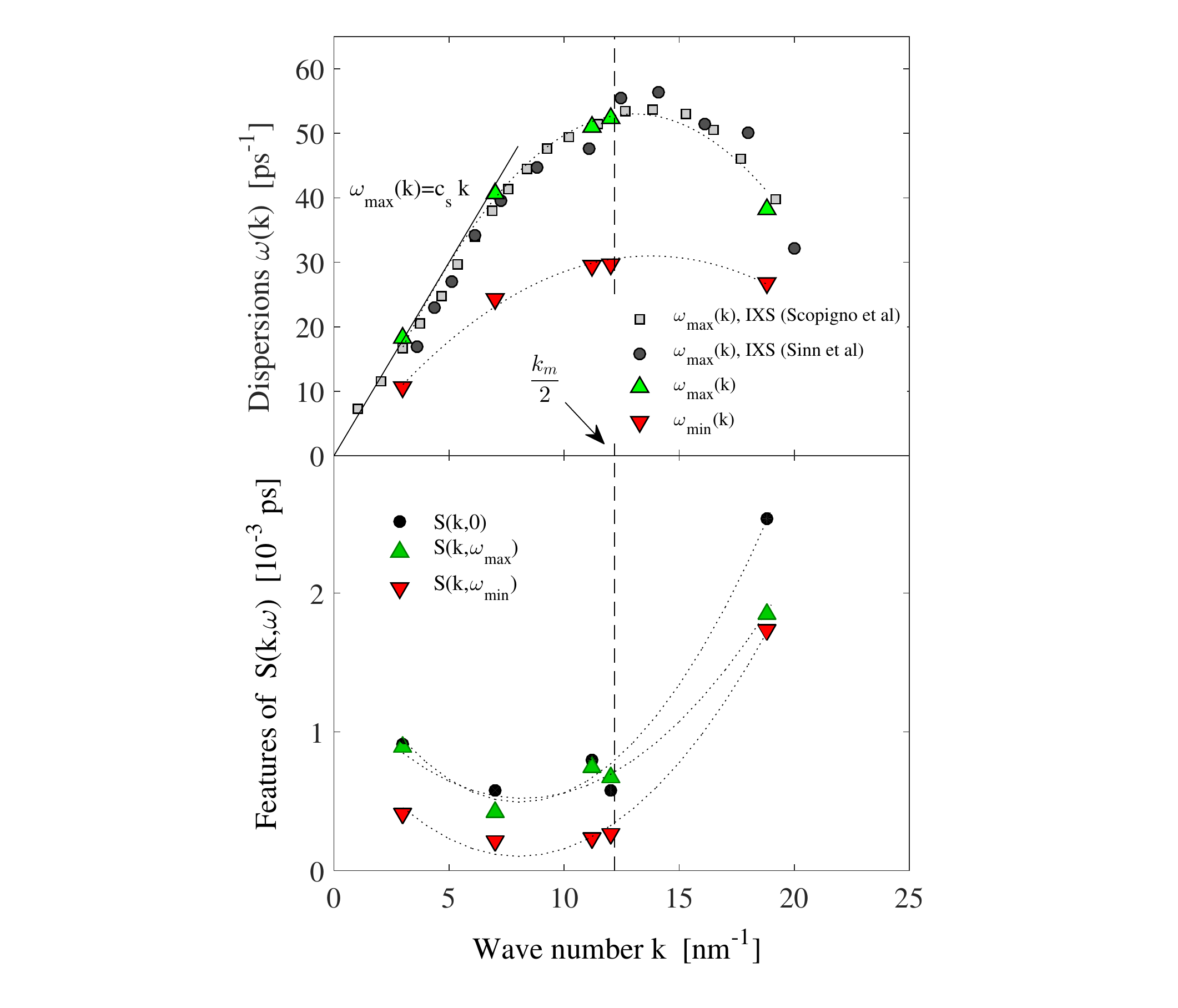}
\caption{Top: Sound dispersion $\omega_{max}(k)$ and wavenumber-dependent frequency
$\omega_{min}(k)$ evaluated from Eqs.~(\ref{eq: side}) and
(\ref{eq: min}); sound dispersions from IXS data of Scopigno \textit{et al.}~\cite{Scopigno_Li1} (full squares) and Sinn \textit{et al.} \cite{Sinn1} (full circles).
Solid line at low
wavenumbers corresponds to hydrodynamic frequency of the
Mandelshtam-Brillouin doublet $\omega_c(k) = c_s k$ with the
adiabatic sound velocity $c_s \simeq 6000$~m/s. Bottom: Wavenumber
dependent spectral features:  zeroth-frequency spectral component
$S(k,\omega=0)$, high-frequency component $S(k,\omega_{max})$,
scattering intensity at the minimum $S(k,\omega_{min})$. Dashed vertical line
corresponds to the border of the first pseudo-Brillouin zone.}\label{fig: parameters}
\end{figure}

In Fig.~\ref{fig: dsf_curr},  the classical dynamic structure
factor $S(k,\omega)$ and the longitudinal current correlation
function $C_L(k,\omega)$ computed on the basis of relations
(\ref{eq: Basic_a}) and (\ref{eq: relation_with_dsf}) are shown.
These results are presented for the same wavenumbers as for the
scattering intensity $I(k,\omega)$ in Fig.~\ref{fig: dsf}. To make
clear correspondence between $S(k,\omega)$ and $C_L(k,\omega)$,
the spectra of $C_L(k,\omega)$ are scaled onto the coefficient
$\alpha(k)$. As seen, the high-frequency peak of $S(k,\omega)$ and
the peak of $C_L(k,\omega)$ are very close, and both the dispersion curves
$\omega_{max}(k)$ and $\omega_c(k)$ have to be practically
identical for the considered wavenumber range~\cite{Brazhkin}. Further, there is
no gap between the central component $S(k,\omega=0)$ and the
high-frequency component $S(k,\omega_{max})$ as it usually have to
be for $S(k,\omega)$ in the hydrodynamic limit (at $k \to
0$)~\cite{Egelstaff1}. So, the lineshape of the dynamic structure
factor $S(k,\omega)$ at the finite $k$ represents the result of
complicated mixing of relaxing and propagative modes. Hence, simple
separation of the density fluctuations into two types associated
with mechanical processes and with thermal processes is not
possible.

The frequency parameters evaluated by means of the two methods are shown in
Fig.~\ref{fig: Deltas}. The wavenumber dependence of these
parameters, $\Delta_1^2(k)$, $\Delta_2^2(k)$, $\Delta_3^2(k)$ and
$\Delta_4^2(k)$, are smooth and similar.  The parameters increase
with increasing wavenumber and saturate at $k \simeq k/k_m
= 12.2$~nm$^{-1}$.  Then, the frequency parameters start to
decrease with further increase of the wavenumber $k$. In fact, the
$k$-dependence of the frequency parameters is similar to the
dispersion curve $\omega_{max}(k)$ of the acoustic-like excitations in
simple liquid~\cite{Pilgrim_review} [compare with Fig.~\ref{fig:
parameters} (top)]. Further, although the parameters $\Delta_1^2(k)$ and
$\Delta_2^2(k)$ take very close values, nevertheless, the
correspondence $\Delta_{j+1}^2(k) \geq \Delta_{j}^2(k)$ for the
frequency parameters $\Delta_j^2(k)$, $j=1,\;2,\;3,\;4$, still
occurs at the considered  wavenumbers. The first three frequency
parameters take the values corresponding to the square $THz$
frequency scale (or to the inverse square $ps$ timescale). Therefore, these correspond to the typical frequencies of local density fluctuations in a liquid at microscopic spatial scales. The frequency parameter $\Delta_4^2(k)$ takes the largest values in comparison with $\Delta_1^2(k)$, $\Delta_2^2(k)$ and $\Delta_3^2(k)$. In fact, the parameter $\Delta_4^2(k)$ defines the upper limit of the frequency domain associated with the density fluctuations. For example, let us consider the
dynamic structure factor $S(k,\omega)$ at $k=3.0$~nm$^{-1}$ [see
the bottom inset in Fig.~\ref{fig: dsf_curr}]. Here, the
scattering function is completely damped at the frequencies
$\omega \geq 30$~ps$^{-1}$. The first three frequency parameters
are $\Delta_1^2=183$~ps$^{-2}$, $\Delta_2^2=210$~ps$^{-2}$ and
$\Delta_3^2=1500$~ps$^{-2}$, whereas the parameter
$\Delta_4^2=177\cdot 10^2$~ps$^{-2}$  is much larger than
$\omega^2 = 900$~ps$^{-2}$. One can see from Fig. \ref{fig: Deltas} that  both the
conditions, $\Delta_4^2(k) \gg \omega^2$ [see also inequality
(\ref{eq: cond})] and $\Delta_4^2(k) \approx \Delta_5^2(k) \approx
\Delta_6^2(k)$ [relation (\ref{eq: time_scales_equality})] are
completely supported by obtained values of the frequency parameters. Moreover, there is the following regularity between the
frequencies $\sqrt{\Delta_1^2(k)}$, $\sqrt{\Delta_2^2(k)}$  and
the spectral features of the dynamic structure factor
$S(k,\omega)$:   both the frequencies $\sqrt{\Delta_1^2(k)}$ and
$\sqrt{\Delta_2^2(k)}$ are located between the minimum and the high-frequency maximum of $S(k,\omega)$, i.e. within
the frequency range $[\omega_{min}(k);\;\omega_{max}(k)]$.

By means of Eqs.~(\ref{eq: sound_vel}), (\ref{eq: D_T}) and (\ref{eq: Gamma}), one can determine the hydrodynamic parameters from the wavenumber dependence of the frequency parameters in the low-$k$ range. We obtain that the sound velocity of liquid lithium at the temperature $T=475$~K is $c_s \simeq 6000 \pm 1200$~m/s, which is larger than the value $4550$~m/s given in Ref.~\cite{Ohse_1985} for liquid lithium at the melting temperature $T_m$. Further, our estimates for the thermal diffusivity yield $D_T \simeq 21.5 \pm 3.5$~nm$^2$/ps. This value is comparable with $19.1$~nm$^2$/ps given for $D_T$ of liquid lithium in Ref.~\cite{Touioukiam}. Finally, we find that the sound attenuation coefficient $\Gamma$ takes the value $\simeq 18.4 \pm 3.5$~nm$^2$/ps.

Since relation~(\ref{eq: Basic_a}) reproduces the lineshape of
the dynamic structure factor $S(k,\omega)$, then all the spectral
features can be also recovered. In Fig.~\ref{fig:
parameters}(top), the frequencies $\omega_{max}(k)$ and $\omega_{min}(k)$ associated with locations of the high-frequency maximum
and minimum of  $S(k,\omega)$ are presented as functions of the
wavenumber $k$. It is necessary to note that  the
dispersion curve $\omega_{max}(k)$ in the low-$k$ limit must be
extrapolated into the hydrodynamic result $\lim_{k \to 0}
\omega_{max}(k) = c_s k$, whereas
$\omega_{min}(k)$  in the low-$k$ asymptotics must be
characterized by a gap, which appears because of the separation of
the Rayleigh and Brillouin components. It is seen in
Fig.~\ref{fig: parameters}(bottom) that the evaluated
components $S(k,\omega=0)$, $S(k,\omega_{max})$ and
$S(k,\omega_{min})$ as functions of the wavenumber are correlated. Namely, the intensities $S(k,0)$ and $S(k,\omega_{min})$ have similar $k-$dependence.
Further, the Rayleigh component $S(k,0)$ and the
Brillouin component $S(k,\omega_{max})$ take practically the
same values for the wavenumbers $k\leq k_m/2$ and yield the ratio
$S(k,0)/S(k,\omega_{max})\simeq 1$.

\section{Concluding remarks \label{sec: concl}}

In liquid, overall wavenumber range available for inelastic
scattering experiments with photons, neutrons and/or X-rays
includes three ranges: the hydrodynamic range (limit), the broad
meso-microscopic range and the free-particle dynamics range
(limit). An extension of hydrodynamic theory is possible up to such
finite wavenumbers,
where the hydrodynamic quantities -- the viscosity, the thermal
conductivity, the sound absorption and others -- still have a
physical meaning. Let us consider the extended wavenumber range,
which includes the wavenumbers corresponding to the hydrodynamic
limit and the wavenumbers associated with mesoscopic-microscopic
scales. According to the ideas of the generalized hydrodynamics, the viscosity
must be transformed into a $k$-dependent parameter when we
consider the wavenumbers corresponding to the mesoscopic range.
Consequently, at microscopical scales comparable with the size of
the second or third pseudo-coordination shells, this macroscopical
parameter (the viscosity) must be transformed into a
characteristic that takes into account the cumulative result of
interparticle interactions within the range. On the other hand,
let the wavenumber $k$ decreases from large values associated with
the limit of a free moving particle to lower values corresponding
to mesoscopic sizes.   Here, with decreasing wavenumber, such the
characteristic as the thermal velocity of a free moving particle
is converted to the phase velocity of the particles enclosed in
appropriate spatial range $\ell \sim 2 \pi/k$. 
Thereofore, the theoretical methodology relevant to describe the
mesoscopic-microscopic dynamics in liquids has to employ such the
quantities as the particle distributions, the interaction energy
between the particles and the frequency characteristics of vibration
dynamics of a particle with respect to its neighbors (similar
to the Einstein frequency).

The theoretical scheme presented in this study exploits the
spectral moments of the dynamic structure factor $S(k,\omega)$ and
does not apply the assumptions about how many relaxing and
propagating modes have to be taken into account. It is important that the
spectral moments are determined by the interparticle potential and
the structural characteristics. On the other hand, the spectral
moments define the sum rules and cannot be considered as arbitrary
model parameters. That is important, the presented theoretical
scheme satisfies all the sum rules of the scattering law.
Moreover, the theoretical relation obtained for the dynamic
structure factor $S(k,\omega)$ transforms approximatively at the
certain conditions associated with convergence of the high-order
frequency parameters to well known hydrodynamic result for
$S(k,\omega)$; as well as the theory yields the correct scenario
for the high-$k$ limit.

The dynamic structure factor is the characteristic of the longitudinal collective dynamics in liquids~\cite{Egelstaff1}. The scattering spectra can also contain marks of the transverse collective excitations with the frequencies lower than $\omega_c(k)=c_s k$~\cite{Giordano_2010}. However, for the case of an equilibrium simple liquid, the intensity of these excitations is very small and it is barely noticeable in the scattering spectra. In particular, this is evidenced by results of Fig. 2 in Ref.~\cite{Hosokawa3}, where representation of the scattering spectra in terms of the corresponding longitudinal and transverse components for liquid Fe, Cu and Zn is given. Thus, it is reasonable to overlook the secondary transverse excitations in the analysis of the longitudinal dynamics. Further, according to the theoretical scheme presented in this study, the spectral density of any dynamical variable relevant to the structural relaxation in a liquid can be expressed in terms of its spectral (frequency) moments. Therefore, the presented theory can be directly extended to the case of the transverse collective dynamics.

Alignment of the high-order frequency parameters does not
contradict to outcomes the linearized hydrodynamics.
In fact, the alignment is identical to the assumption about equality of
the TCF's $M_3(k,t)$ and $M_4(k,t)$. This allows one to define
self-consistently the intermediate scattering function $F(k,t)$
and all other TCF's: $M_1(k,t)$, $M_2(k,t)$ and $M_3(k,t)$. In
such a view, the presented theoretical scheme is similar in its
construction to the mode-coupling theories adapted for ergodic systems~\cite{Gotze_book_2009}, where
the memory function $M_2(k,t)$ is expressed in terms of the
intermediate scattering function $F(k,t)$. It should be
noted that  Reichman and Charbonneau [in
Ref.~\cite{Reichman_closure} (discussion on p.~21)] pointed to a
possible extension of the mode-coupling theory, involving the
shift of the ``mode-coupling closure'' to the high-order memory
function $M_4(k,t)$. This is directly corresponds to the
theoretical scheme presented in this study.

In this study, we applied the theoretical approach to determine the dynamic structure factor and the scattering intensity at some selected wavenumbers  for the case of liquid lithium. We chose this system to test our theory for the following reasons. As asserted in Ref.~\cite{Tankeshwar_2007}, the collective microscopic dynamics of this liquid can be more complicated than the dynamics of other liquid metals. Therefore, a microscopic theory may be experiencing difficulties in its application to liquid lithium. On the other hand, there are available the experimental data of the high quality for the microscopic structure and collective dynamics of liquid lithium~\cite{Salmon,Sinn1,Scopigno_Li1}. We found that the theory is capable of reproducing all the features of the scattering intensity in agreement with IXS data.  Finally, the presented theory  utilizes the particle interaction potential and structural characteristics as input parameters. Therefore, this theory can be applied directly to the case of any simple fluid.

\ack We thank T. Scopigno for providing us with experimental IXS
data and A.G. Novikov, V.V. Brazhkin and V.N. Ryzhov  for useful
discussions. We also grateful to L.E. Gonz\'{a}lez for his advice
on the lithium pseudopotential and to R.M. Khusnutdinoff for his
help with molecular dynamics simulations. 
The reported study was supported in part by RFBR according to the research project No. 18-02-00407.
Authors are thankful to the Ministry of
Education and Science of the Russia Federation for supporting the
research in the framework of the state assignment
(3.2166.2017/4.6).

\section*{Appendix: The frequency parameters from configuration data. Details of molecular dynamics simulations}

The wavenumber-dependent frequency parameters $\Delta_1^2(k)$,
$\Delta_2^2(k)$, $\Delta_3^2(k)$, $\ldots$, can be calculated
on the basis of configuration data of the considered many-particle
system. These data must include the coordinates, velocities and
accelerations, i.e. $\{\mathbf{r}_1,\; \mathbf{v}_1,\; \mathbf{a}_1;\; \mathbf{r}_2,\; \mathbf{v}_2,\; \mathbf{a}_2;\; \ldots ; \; \mathbf{r}_N,\; \mathbf{v}_N,\; \mathbf{a}_N\}$,  of all the $N$ particles forming the system, that is
usually available, for example, from molecular dynamics
simulations  with the given interparticle potential. Then, the dynamical variables $A_0(k)$, $A_1(k)$,
$A_2(k)$ \textit{etc.} can be computed on the basis of Eq.~(\ref{eq:
density_fluct}) and recurrent relation~(\ref{eq: variables}),
whereas the frequency parameters can be computed from
Eq.~(\ref{eq: frequency_parameter}).

In particular, we find for the first five dynamical variables the following exact expressions:
\begin{equation}\label{eq_appendix_5}
A_{0}(k,t)\equiv\rho_{k}(t)=\frac{1}{\sqrt{N}}\sum_{n=1}^{N}\left[\cos(\mathbf{k}\cdot \mathbf{r}_{n}(t))+i\sin(\mathbf{k}\cdot \mathbf{r}_{n}(t))\right],
\end{equation}
\begin{equation}\label{eq_appendix_6} A_{1}(k,t)=\frac{1}{\sqrt{N}}\sum_{n=1}^{N}\left[\Bigg(\mathbf{k}\cdot\mathbf{v}_{n}(t)\Bigg)\Bigg\{-\sin(\mathbf{k}\cdot \mathbf{r}_{n}(t))+i\cos(\mathbf{k}\cdot \mathbf{r}_{n}(t))\Bigg\}\right],
\end{equation}
\begin{eqnarray}\label{eq_appendix_7}
A_{2}(k,t) &=& \frac{1}{\sqrt{N}}\sum_{n=1}^{N}\left[\Bigg(\mathbf{k}\cdot\mathbf{a}_{n}(t)\Bigg)\Bigg\{-\sin(\mathbf{k}\cdot \mathbf{r}_{n}(t))+i\cos(\mathbf{k}\cdot \mathbf{r}_{n}(t))\Bigg\} \right . \nonumber \\ & & \left .  + \Bigg(\mathbf{k}\cdot\mathbf{v}_{n}(t)\Bigg)^{2}\Bigg\{-\cos(\mathbf{k}\cdot \mathbf{r}_{n}(t))-i\sin(\mathbf{k}\cdot \mathbf{r}_{n}(t))\Bigg\}\right] \nonumber \\ & + & \Delta_{1}^{2}(k)A_{0}(k,t),
\end{eqnarray}
\begin{eqnarray}\label{eq_appendix_8}
A_{3}(k,t) &=& \frac{1}{\sqrt{N}}\sum_{n=1}^{N}\left[\Bigg(\mathbf{k}\cdot\frac{d\mathbf{a}_{n}(t)}{dt}\Bigg)\Bigg\{-\sin(\mathbf{k}\cdot \mathbf{r}_{n}(t))+i\cos(\mathbf{k}\cdot \mathbf{r}_{n}(t))\Bigg\} \right . \nonumber \\ & & \left .  +  \Bigg(\mathbf{k}\cdot\mathbf{v}_{n}(t)\Bigg)^{3}\Bigg\{\sin(\mathbf{k}\cdot \mathbf{r}_{n}(t))-i\cos(\mathbf{k}\cdot \mathbf{r}_{n}(t))\Bigg\} \right . \nonumber \\ & & \left .  + 3\Bigg(\mathbf{k}\cdot\mathbf{v}_{n}(t)\Bigg)\Bigg(\mathbf{k}\cdot\mathbf{a}_{n}(t)\Bigg)\Bigg\{-\cos(\mathbf{k}\cdot \mathbf{r}_{n}(t))-i\sin(\mathbf{k}\cdot \mathbf{r}_{n}(t))\Bigg\}\right] \nonumber \\ & + &
\left[\Delta_{1}^{2}(k)+\Delta_{2}^{2}(k)\right]A_{1}(k,t)
\end{eqnarray}
and
\begin{eqnarray}\label{eq_appendix_9}
A_{4}(k,t) &=&
\frac{1}{\sqrt{N}}\sum_{n=1}^{N}\left[\Bigg(\mathbf{k}\cdot\frac{d^{2}\mathbf{a}_{n}(t)}{dt^{2}}\Bigg)\Bigg\{-\sin(\mathbf{k}\cdot \mathbf{r}_{n}(t))+i\cos(\mathbf{k}\cdot \mathbf{r}_{n}(t))\Bigg\} \right . \nonumber \\ & & \left .  + 4\Bigg(\mathbf{k}\cdot\mathbf{v}_{n}(t)\Bigg)\Bigg(\mathbf{k}\cdot\frac{d\mathbf{a}_{n}(t)}{dt}\Bigg)\Bigg\{-\cos(\mathbf{k}\cdot \mathbf{r}_{n}(t))-i\sin(\mathbf{k}\cdot\mathbf{r}_{n}(t))\Bigg\} \right . \nonumber \\ & & \left .  + 3\Bigg(\mathbf{k}\cdot\mathbf{a}_{n}(t)\Bigg)^{2}\Bigg\{-\cos(\mathbf{k}\cdot \mathbf{r}_{n}(t))-i\sin(\mathbf{k}\cdot \mathbf{r}_{n}(t))\Bigg\} \right . \nonumber \\ & & \left .  + 6\Bigg(\mathbf{k}\cdot\mathbf{v}_{n}(t)\Bigg)^{2}\Bigg(\mathbf{k}\cdot\mathbf{a}_{n}(t)\Bigg)\Bigg\{\sin(\mathbf{k}\cdot \mathbf{r}_{n}(t))-i\cos(\mathbf{k}\cdot \mathbf{r}_{n}(t))\Bigg\} \right . \nonumber \\ & & \left .  + \Bigg(\mathbf{k}\cdot\mathbf{v}_{n}(t)\Bigg)^{4}\Bigg\{\cos(\mathbf{k}\cdot \mathbf{r}_{n}(t))+i\sin(\mathbf{k}\cdot \mathbf{r}_{n}(t))\Bigg\}\right] \nonumber \\ & + & \Bigg(\Delta_{1}^{2}(k)+\Delta_{2}^{2}(k)+\Delta_{3}^{2}(k)\Bigg)\Bigg(A_{2}(k,t)-\Delta_{1}^{2}(k)A_{0}(k,t)\Bigg)\nonumber \\ & + & \Delta_{1}^{2}(k)\Delta_{3}^{2}(k)A_{0}(k,t).
\end{eqnarray}

The time derivatives $d\mathbf{a}_{n}(t)/dt$ and $d^{2}\mathbf{a}_{n}(t)/dt^{2}$ in Eqs.~(\ref{eq_appendix_8}) and (\ref{eq_appendix_9}) can be determined numerically  by the finite difference method. Namely, one can apply the next simple scheme:
\begin{equation}\label{eq_appendix_10}
\frac{d\mathbf{a}_{n}(t_{i+1})}{dt}=\lim_{\Delta t\rightarrow0}\frac{\mathbf{a}_{n}(t_{i+1})-\mathbf{a}_{n}(t_{i})}{\Delta t},
\end{equation}
\begin{equation}\label{eq_appendix_11}
\frac{d^{2}\mathbf{a}_{n}(t_{i+2})}{dt^{2}}=\lim_{\Delta t\rightarrow0}\frac{\mathbf{a}_{n}(t_{i+2})-2\mathbf{a}_{n}(t_{i+1})+\mathbf{a}_{n}(t_{i})}{\Delta t},
\end{equation}
with the simulation time step $\Delta t=1\,$fs and $i=1,\,2,\,3,\,...$. Further, by applying relation~(\ref{eq: variables}), one can obtain expressions for the dynamical variables of a higher order. Then, Eq.~(\ref{eq: frequency_parameter}) can be rewritten in the form adapted for the numerical estimation of the frequency parameters:
\begin{eqnarray}\label{eq_appendix_1}
\Delta_{j+1}^{2}(k)=\frac{\left\langle\left\{\mathrm{Re} [ A_{j+1}(k,t)] \right \}^{2}+\left\{\mathrm{Im} [A_{j+1}(k,t)]\right \}^{2}\right\rangle}{\left\langle\left \{ \mathrm{Re} [A_{j}(k,t)] \right \}^{2}+\left \{ \mathrm{Im} [A_{j}(k,t)] \right \}^{2}\right\rangle}, \\
j=1,\;2,\;3,\ldots . \nonumber
\end{eqnarray}
\begin{figure}[ht]
\centering
\includegraphics[width=1.0\linewidth]{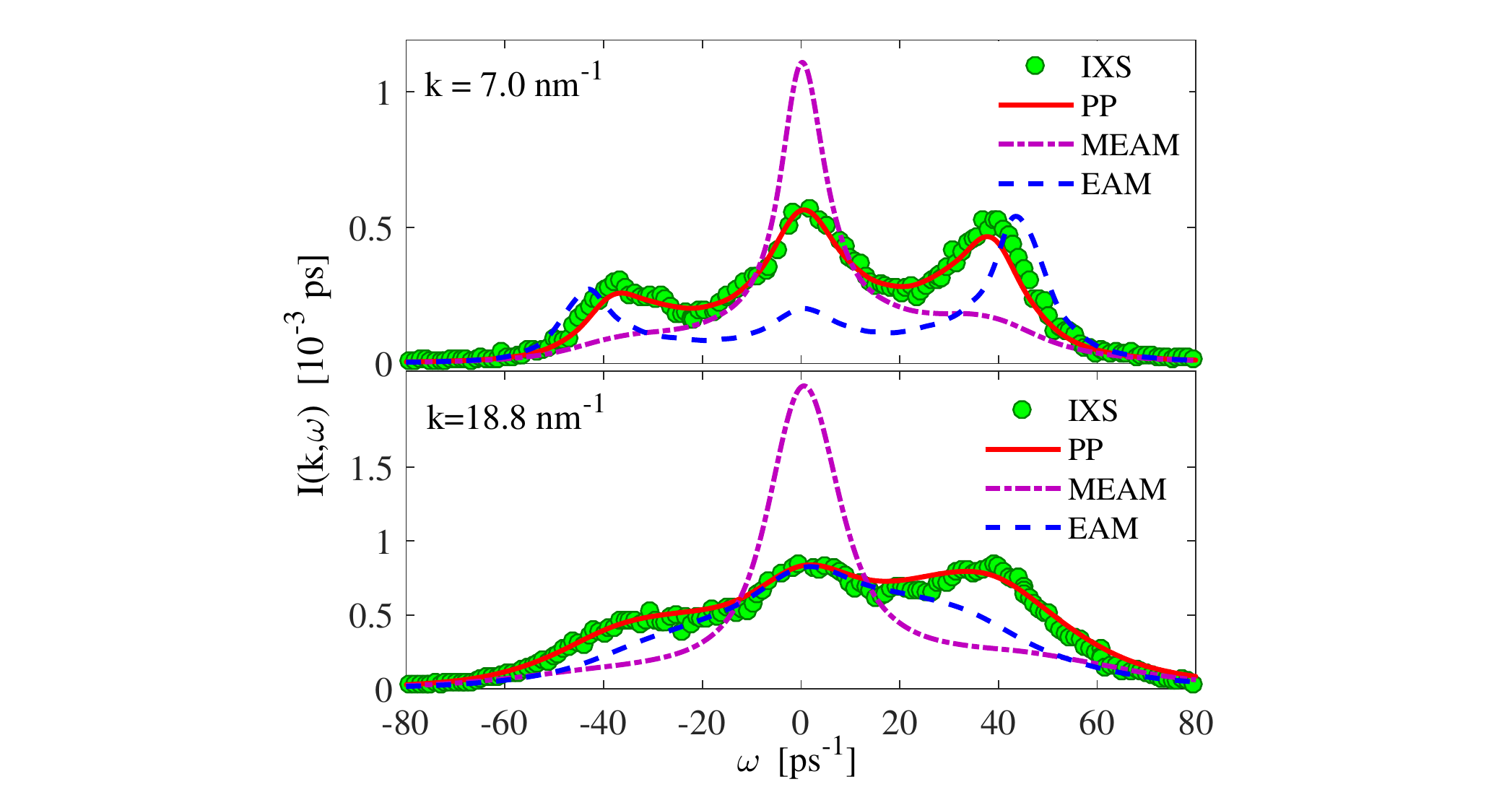}
\caption{Scattering intensity of liquid lithium at the temperature $T=475$~K and at the selected wavenumbers. Comparison experimental IXS data~\cite{Scopigno_Li1} and results of MD simulations with the model pseudopotential~\cite{Gonz_2001}, with the EAM potential~\cite{Belashenko} and with the MEAM potential~\cite{Baskes}.}
\label{fig: Spectra_MD}
\end{figure}
For the case of the isotropic
equilibrium system, the ensemble average in Eq.~(\ref{eq_appendix_1}) is realized by means of the averaging over
configurations at different time moments and averaging over
different directions of the wavevector $\mathbf{k}$ at the fixed
 \[
 k=|\mathbf{k}|=\sqrt{k_{x}^{2}+k_{y}^{2}+k_{z}^{2}}
 \]
 within the given geometry of the simulation box.

\vskip 0.5cm

Note that we have verified the various model potentials suggested for liquid lithium. It is remarkable that the pseudopotential proposed by Gonzalez \textit{et al.} yields very good agreement with X-ray diffraction data for the structure properties and  excellent agreement with IXS data for the dynamical properties~\cite{Khusn_2018}. In particular, the scattering intensity $I(k,\omega)$  calculated with this pseudopotential reproduces the experimental data with a higher accuracy than the scattering intensity obtained with the EAM~\cite{Belashenko} and MEAM potentials~\cite{Baskes} (see Fig.~\ref{fig: Spectra_MD}). Therefore, we have used this pseudopotential to generate the configuration data and, then, to estimate the frequency parameters.

Thus, the first six frequency parameters $\Delta_1^2(k)$,
$\Delta_2^2(k)$, \ldots, $\Delta_6^2(k)$ of the liquid lithium for
the considered wavenumber range were calculated on the basis of
 molecular dynamics simulation data for the system of
$N=16\;000$ particles interacted through the effective
pseudopotential suggested by Gonzalez \textit{et al.}~\cite{Gonz_2001}. It was
employed the cubic simulation box ($L_{x}=L_{y}=L_{z}=6.98\,$nm)
with the periodic boundary conditions in all the directions. The
values of the frequency parameters were also averaged over data of
independent simulations runs. The obtained values of these parameters are presented in Fig.~\ref{fig: Deltas}.

\end{document}